%% file: bare_jrnl.tex
\documentclass[journal]{IEEEtran}

\usepackage{textcomp}
\usepackage{xcolor}
\usepackage{booktabs}
\usepackage{multirow}
\usepackage{adjustbox}
\usepackage{array}
\usepackage{color}
\usepackage{epsf}
\usepackage{times}
\usepackage{epsfig}
\usepackage{graphicx}
\usepackage{epstopdf}
\usepackage{hyperref}
\usepackage{cite}
\usepackage{amsmath}
\usepackage{amssymb}
\usepackage{amsxtra}
\usepackage{amsthm}
\usepackage{bbm}
\usepackage{caption}
\usepackage{algorithm}

\usepackage{subcaption}
\usepackage{authblk}
\usepackage{bbm}
\usepackage{comment}
\usepackage{soul}

\usepackage{algpseudocode}
\usepackage{algorithm}
\usepackage[mathscr]{euscript}

\input{Figures/defs_tikzpgf}
\usetikzlibrary{petri}
\usetikzlibrary{shapes}
\usetikzlibrary{positioning,arrows,patterns}
\usepackage{scalefnt}
\usetikzlibrary{calc,decorations.markings}
\pgfplotsset{compat=1.10}
\usetikzlibrary{arrows.meta}
\usepackage[columnwise,switch,mathlines]{lineno} 

\usepackage{pgfplots}
\pgfplotsset{compat=newest}
\usepgfplotslibrary{groupplots}

\usepackage{xcolor,cite,etoolbox}
\makeatletter
\pretocmd\@bibitem{\color{black}\csname keycolor#1\endcsname}{}{\fail}
\newcommand\citecolor[1]{\@namedef{keycolor#1}{\color{blue}}}
\makeatother

\usepackage{cite}

\IEEEoverridecommandlockouts
\newtheorem{theorem}{Theorem}
\newtheorem{lemma}{Lemma}

\begin{document}
\title{Characterization of the Global Bias Problem in Aerial Federated Learning}
\author{Ruslan Zhagypar,~\IEEEmembership{Student Member,~IEEE,}
        Nour~Kouzayha,~\IEEEmembership{Member,~IEEE,}
		Hesham~ElSawy,~\IEEEmembership{Senior Member,~IEEE,}
		Hayssam~Dahrouj,~\IEEEmembership{Senior Member,~IEEE,}
		and~Tareq Y. Al-Naffouri,~\IEEEmembership{Senior Member,~IEEE}
		\addvspace{-2\baselineskip}
 		\thanks{R. Zhagypar, N. Kouzayha, and T. Y. Al-Naffouri are with King Abdullah University of Science and Technology, Thuwal, Saudi Arabia (e-mail: ruslan.zhagypar@kaust.edu.sa; nour.kouzayha@kaust.edu.sa; tareq.alnaffouri@kaust.edu.sa).}
 		\thanks{H. ElSawy is with the School of Computing, Queen's University, ON, Canada (e-mail: hesham.elsawy@queensu.ca).}
		\thanks{H. Dahrouj is with the Department of Electrical Engineering, University of Sharjah, United Arab Emirates (e-mail: hayssam.dahrouj@gmail.com).}\vspace{0pt}
	}
\maketitle
\IEEEpeerreviewmaketitle
\begin{abstract}
   Unmanned aerial vehicles (UAVs) mobility enables flexible and customized federated learning (FL) at the network edge. However, the underlying uncertainties in the aerial-terrestrial wireless channel may lead to a biased FL model. In particular, the distribution of the global model and the aggregation of the local updates within the FL learning rounds at the UAVs are governed by the reliability of the wireless channel. This creates an undesirable bias towards the training data of ground devices with better channel conditions, and vice versa. 
   This paper characterizes the global bias problem of aerial FL in large-scale UAV networks. To this end, the paper proposes a channel-aware distribution and aggregation scheme to enforce equal contribution from all devices in the FL training as a means to resolve the global bias problem. We demonstrate the convergence of the proposed method by experimenting with the MNIST dataset and show its superiority compared to existing methods. The obtained results enable system parameter tuning to relieve the impact of the aerial channel deficiency on the FL convergence rate.
\end{abstract}
\vspace{-0.4cm}
\section{Introduction}
\IEEEPARstart{T}he development of the sixth generation (6G) of wireless networks enables the joint use of unmanned aerial vehicles (UAVs) and federated learning (FL), so as to harness the prospects of learning on the edge via aerial platforms. The potential of UAVs coupled with the FL algorithm has been recognized in~\cite{wang2021learning}\cite{lim2021uav}. The FL algorithm facilitates the progress of the Internet of drones and drones-as-a-service~\cite{lim2021towards}.
Moreover, the FL-based edge computing in a UAV network enhances client quality-of-service and maintains privacy~\cite{abreha2022federated}. However, the high communication cost between an FL aggregator and learners remains the main obstacle to practical implementation. Thus, the work in~\cite{chu2022design} reduces this cost by clients clustering and transfer learning at the price of smaller accuracy. Authors of \cite{mestoukirdi2022uav} advocate the role of UAVs in FL over  terrestrial base stations (BS) due to UAVs mobility, which allows for customized learning from a specific group of devices. Not only do UAVs offer more extensive coverage, but their on demand deployment also makes them a cost-effective solution~\cite{kouzayha2021analysis}.

The basic principle of the FL algorithm is to send the local parameters to an aggregator and receive back the updated global model parameters. Since these transmissions occur over wireless channels, the FL operation becomes eventually governed by the conditions of the channel. Due to the unreliable and stochastic nature of the wireless channel, some local updates may be lost after transmission at the aggregator. Furthermore, the global model broadcasting might not reach the scheduled devices on account of UAVs' limited energy resources to compensate for channel losses. These events create an undesirable bias towards devices with more stable wireless connectivity, which drifts the global FL model away from optimality. As a result, the model would suffer from inefficient learning in terms of lower accuracy and slower convergence. To this end, this paper considers a single-tier network consisting of UAVs that act as aggregators and ground devices that execute local FL learning. The paper then  characterizes the global bias problem by deriving the UAV’s download and clients’ upload success probabilities, and by integrating such metrics in the aggregation step of the FL process.

In the existing literature, several studies attempt to account for the wireless channel's impact on the convergence of FL algorithms in terrestrial networks~\cite{amiri2020federated}\cite{dinh2021federated}. To this end, stochastic geometry is utilized as the main mathematical framework to provide analytical formulations of key performance metrics that assist the FL algorithms. For instance, in \cite{salehi2021federated}, the authors calculate the client update success probability and highlight its impact along with resource constraints on the learning performance. Theoretical analysis made in \cite{lin2022deploying} and \cite{yang2020scheduling} show that the spatial convergence of FL depends on the number of devices with high uplink (UL) success probability. Authors of \cite{xia2022stochastic} apply the framework to explore the effect of dynamic signal-to-interference-plus-noise ratio (SINR) threshold on learning.

In view of prior work, most studies assume a guaranteed downlink (DL) transmission because of sufficient power resources of the terrestrial BSs. However, the perfect global model reception is no longer true in the setting of aerial wireless networks due to UAVs' limited energy capacity. Thus, despite the high potential of UAVs in flexible edge intelligence, no study specifically solves the global bias problem of FL in the setting of UAV-assisted wireless networks. 
To the best of the authors' knowledge, this is the first work that bridges this gap by optimizing a reliable FL algorithm for aerial networks.

In this work, we propose a UAV-assisted FL algorithm in which UAVs provide an intermediate model aggregation from the sky and communicate with ground devices through unreliable wireless channels. Inspired by the results in~\cite{salehi2021federated}, we jointly characterize the UAV's download and clients' upload success probabilities and integrate them in the aggregation step of the FL to counter the mentioned bias problem.
The main contributions of this work can, therefore, be summarized as follows: 
1) Developing an analytical framework for modeling and analysis of aerial FL in large-scale UAV networks. The analytical model is used to derive a tractable expression for the joint DL and UL success probability as a function of the different system parameters.
2) Unbiasing the FL algorithm by integrating the joint success probability in the aggregation step. By experimenting on the MNIST dataset, we validate the convergence of the proposed FL and show its superiority over the existing FL algorithms.
3) Based on the proposed framework, we investigate the impact of different system parameters including the height of UAVs and the effects of different environments. The devised insights can be used to tune the system parameters to achieve optimal learning performance.

\begin{figure}[t]
     \centering
      {\includegraphics[width=0.95\linewidth]{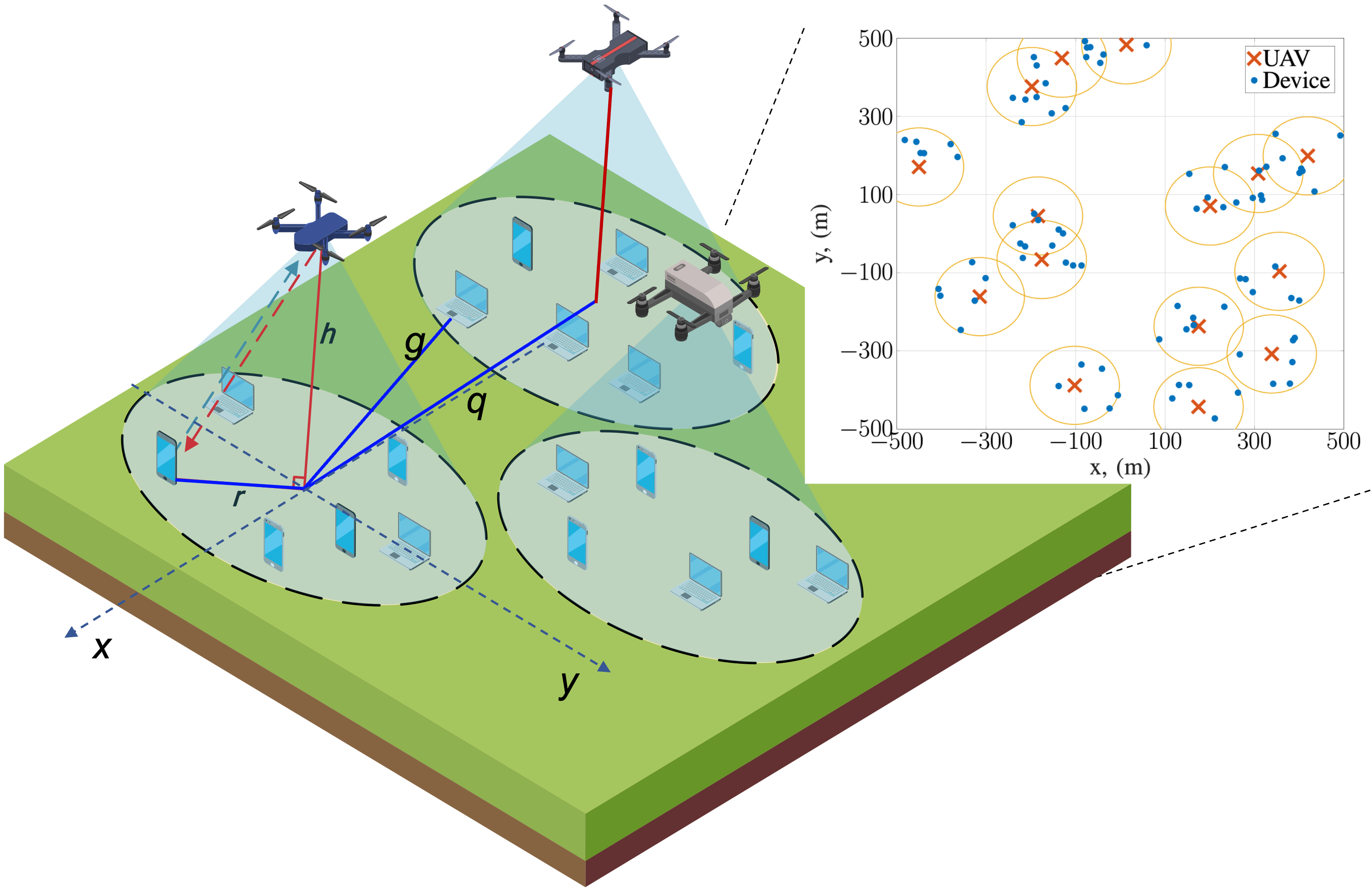}
      }
     \setlength{\belowcaptionskip}{-12pt}
     \captionsetup{font=footnotesize}
     \vspace{-0.1cm}
     \caption{Illustration of the system model and an example of MCP with $N=6$ in a cluster of radius $R=100$~m.}
    \label{fig:system_model1}
\end{figure}

\section{System Model}
We consider a single-tier network consisting of UAVs that act as aggregators and ground devices that execute local FL learning, as illustrated in Fig.~\ref{fig:system_model1}. As devices performing the FL with the same UAV are physically close to each other, we utilize a Matern Cluster Process (MCP) to model the network. This model also represents the case where a UAV flies up to a cluster of devices and hovers over it to conduct an FL training on their datasets. The UAVs are considered the parent points and their locations follow a homogeneous Poisson Point Process (PPP) $\Phi$ with intensity $\lambda$. Around each UAV located at $u\in\Phi$, a set $D_u$ of $N$ devices are independently and identically distributed (iid) forming a cluster of radius $R$. 
An example of the MCP model with a cluster radius of $R=100$~m and $N=6$ is illustrated in the foreground of Fig.~\ref{fig:system_model1}. Each UAV orchestrates a separate FL process by aggregating the local updates from the devices within its cluster. 


As the aerial links are affected with the blockages in the environment and since the devices are considered at the ground level, we use the approximation of the line-of-sight (LOS) probability proposed in \cite{al2014optimal}. Specifically, the probability of having a LOS link between the UAV and a device $k$ located at a horizontal distance $r_k$ is given by:
\begin{equation}
P_L(r_k) = \frac{1}{1+a\exp{(-b[\frac{180}{\pi}\arctan{(h/r_k)}-a])}},
\end{equation}
where $a$ and $b$ are constants that depend on the environment and are given
in~\cite[Table I]{bor2016efficient}, and $h$ is the elevation height of the UAV. The non line-of-sight (NLOS) probability is thus given as $P_N(r_k) = 1 - P_L(r_k)$. As a result, the set of UAVs is divided into two subsets with intensities $P_L(r_k) \lambda$ and $P_N(r_k) \lambda$. We assume different path loss exponents and fading parameters for LOS and NLOS transmissions.

We assume that both the UAVs and devices can perform directional beamforming to compensate for propagation path loss. Hence a sectored antenna model is assumed at the UAVs and ground devices, where $M_s$ and $m_s$ are the main lobe gain and side lobe gain, respectively, and $s\in \{u,d\}$ is the index that denotes the UAV and the device, respectively. We assume that the UAV and devices can steer their antennas to maximize the directionality gain. Thus, the gain on the desired link achieves the maximum value of $G_0=M_uM_d$. On the other side, the beam directions of the interfering links
are uniformly distributed over $[0, 2\pi)$. Thus, we can formulate the directionality gain of an interfering link as shown in Table~\ref{tab:my-table2} where the probabilities are functions of the main lobe beamwidth $\theta_s$ for $s\in \{u,d\}$.


\begin{table}[]
\centering
\caption{Four antenna patterns}
\label{tab:my-table2}
\resizebox{\columnwidth}{!}{%
\begin{tabular}{|c|c|c|c|c|}
\hline
$i$    & \textbf{1} & \textbf{2} & \textbf{3} & \textbf{4} \\ \hline
$G_i$ &$M_uM_d$    &$M_um_d$    &$m_uM_d$    & $m_um_d$  \\ \hline
$p_i$ & $(\frac{\theta_u}{2\pi})(\frac{\theta_d}{2\pi})$  & $(\frac{\theta_u}{2\pi})(1-\frac{\theta_d}{2\pi})$      &$(1-\frac{\theta_u}{2\pi})(\frac{\theta_d}{2\pi})  $    & $ (1-\frac{\theta_u}{2\pi})(1-\frac{\theta_d}{2\pi})$  \\ \hline
\end{tabular}%
}
\vspace{-4mm}
\end{table}

We account for the impact of the distance-dependent path-loss and small-scale fading modeled by the Nakagami-$m$ distribution. Universal frequency reuse is used across different clusters and $M$ resource blocks (RBs) are available at each UAV. Due to the scarcity of resources ($M\le N$), we apply random resource scheduling without replacement for devices in each cluster. The scheduling also eliminates intra-cluster interference. We utilize the SINR as a metric  to characterize the UL channel used for transmitting local parameters from devices to UAVs. Thus, the UL SINR is given by:
\begin{equation}
    \text{SINR}^{\text{UL}}_z = \frac{P_d G_0 |h_k|_z^2 (r_k^2+h^2)^{-\frac{\alpha_z}{2}}}{I_{\text{UL}}+n_0^2},
    \vspace{-0.1cm}
\end{equation}
where $z\in\{{L, N}\}$ denotes the LOS and NLOS links, $\alpha$ is the path-loss coefficient, $P_d$ is the transmission power of a device, $G_0$ is the directionality gain of the desired link, $|h_k|^2_z$ is the power of the normalized small-scale Nakagami-$m$ fading, $r_k$ is the distance between projections of a selected UAV and transmitting device $k$, $n_0^2$ is the noise power, and $I_{\text{UL}}$ represents the interference from inter-cluster devices.

The DL channel used for transmitting the global parameters from UAVs to devices is also prone to errors due to interference, propagation path-loss, and noise. The SINR is also utilized to characterize the DL channel, which is given by
\begin{equation}
    \text{SINR}^{\text{DL}}_z = \frac{P_u G_0 |h_k|_z^2 (r_k^2+h^2)^{-\frac{\alpha_z}{2}}}{I_{\text{DL}}+n_0^2},
    \vspace{-0.1cm}
\end{equation}
where $P_u$ is the transmission power of a UAV, and $I_\text{DL}$ is the interference from other UAVs.

Fig.~\ref{fig:system_model1} demonstrates the projections of necessary distances for the characterization of inter-cluster interference. The UAV of the typical cluster, $u_0 \in \Phi$, is positioned right above the origin at an altitude $h$. Thus, $g = ||d||, d\in D_{u\backslash u_0}$ represents the Euclidean distance to a device in an interfering cluster. Further, let  $q=||u||$, where $u\in \Phi\backslash u_0$ denotes the distance to an interfering UAV. The conditional distance $g|q$ distribution for MCP is given as~\cite{yi2019unified}:
\begin{equation}
\begin{aligned}
\ \ \ \ f_Q^{\text{MAT}}(g|q) = \frac{2g}{\pi R^2}\arccos{\left(\frac{g^2+q^2-R^2}{2gq}\right)}\textbf{U}(g-|R-q|)\\
\times \textbf{U}(R+q-g) + \frac{2g}{\pi R^2}\textbf{U}(R-q-g),
\label{condDist}
\end{aligned}
\end{equation}
 with $\textbf{U}(\cdot)$ is the unit step function.
 
\section{FL Algorithm}
In the FL scenario, geographically dispersed devices interact with a central server, located at the UAV, to train a common learning model. For learning a statistical model from the distributed data, the central server aims to solve the following optimization problem:
\begin{equation}
    \min_w \ F(w) = \sum^N_{k=1}p_k F_k(w),
    \label{GlobalF}
\end{equation}
where $w$ is the learning model parameter, $p_k=n_k/n$ represents the weight of device $k$ with $n_k$ samples in its local dataset $\mathscr{D}_k$, and $n$ is the total number of data samples.
$F_k(w)$ denotes the local loss function at device $k$, given by:
\begin{equation}
    F_k(w) = \frac{1}{n_k}\sum_{x \in \mathscr{D}_k} f(w,x),
\end{equation}
where $f(w,x)$ is the point local loss function for data sample $x$. It should be noted that the dataset $\mathscr{D}_k$ is non-iid across different devices. Since the aggregator cannot directly solve (\ref{GlobalF}), an iterative approach should be applied. Hence, we propose the algorithm shown in Algorithm 1, which solves the FL problem in (\ref{GlobalF}) while considering the unreliable DL global model distribution and UL local models aggregation in aerial setting.
The aggregation step is the key feature of Algorithm 1 when compared to existing methods. It is worth noting that the work in \cite{salehi2021federated} assumes perfect DL reception and erroneous UL channel for terrestrial settings. In conventional FedAvg algorithms \cite{mcmahan2017communication}, the global aggregation rule at the server does not consider any channel unreliability and is performed as
\vspace{-0.1cm}
\begin{equation}
\vspace{-0.1cm}
     w_{t+1} = \sum^{N}_{k=1} p_k w^k_{t+1},
\end{equation}
where $w^k_{t}$ is the local model parameter of device $k$ at time $t$. Since in the proposed method we allow for $E$ number of local stochastic gradient descent (SGD) iterations, the value of $w^k_{t}$ takes the following form:
\begin{equation}
w^k_{t}=
    \begin{cases}
        w_0 & \text{if } t=0\\
        w_t & \text{if } t \in \{E, 2E, 3E,...\}\\
        v^{k}_{t} & \text{otherwise},
    \end{cases}
\end{equation}
where the second case denotes the local initialization stages, and $v^{k}_{t}$ is the locally updated model parameter. Note that the local learning in Algorithm 1 is performed in batches of $\xi_t^k$ on device $k$ at time $t$.
\begin{algorithm}[t]
\caption{FL for UAV-assisted wireless networks}
\textbf{Server Executes:}
\begin{algorithmic}
\State initialize $w_0$
\For{each round $t = 0, 1, 2, ...$} 
    \State $S_t$ $\leftarrow$  (random set of $M$ clients out of $N$)
    \State Broadcast $w_t$ to the clients 
        \For{each client $k \in S_t$ in parallel}
            \State $v^{k}_{t+1} \leftarrow$ LocalUpdate($k$, $w_t$)
            \State Calculate the joint success probability $J_k$ using (\ref{UkFormula})
        \EndFor 
    \State $w_{t+1} = w_t +\sum_{k=1}^{N}\sum_{b=1}^{M} \frac{p_k}{q_k J_k} \mathbbm{1}(\text{SINR}_{k, b}^{\text{DL}}>\tau_{\text{DL}},\ \text{SINR}_{k, b}^{\text{UL}}>\tau_{\text{UL}})(v^k_{t+1}-w_t)$ //aggregation step
\EndFor
\end{algorithmic}
\textbf{LocalUpdate($k$, $w_t$)}:
\begin{algorithmic}
\State $\xi_t^k \leftarrow$ (split $\mathscr{D}_k$ into batches)
\For{each local epoch $i$ from 0 to $E-1$}
\State $w^k_{t+i} = w^k_{t} - \eta_{t+i} \nabla F_k(w^k_{t+i},\xi_t^k )$
\EndFor
\State Transmit $w^k_{t+E}$ to the server
\end{algorithmic}
\end{algorithm}

In this paper, we apply random scheduling in the following manner: the UAV selects uniformly $M$ out of $N$ devices without replacement, thereby guaranteeing that each device uses at most one RB. Hence, on average, the UAV allocates a RB to device $k$ with probability $q_k$. Thus, $q_k$ is defined as
\begin{equation}
    q_k = \mathbb{E}\left[\sum^M_{b=1}\mathbbm{1}(k \in S_t(b))\right] = \frac{M}{N},
\end{equation}
where $S_t(b)$ represents a set of devices scheduled at time $t$ for a RB $b$.
The indicator function in Algorithm 1 captures the unreliability of the wireless channel during UL/DL transmission. To counteract the impact of the wireless channel, the weights of local updates are divided by the joint success probability $J_k$ derived in Theorem 1 below. The proof of convergence of the proposed algorithm is similar to the one shown in \cite{salehi2021federated}, and is omitted in this paper due to space limitations.
\begin{theorem}\normalfont
In a UAV-assisted wireless network modeled as an MCP, the joint success probability for device $k$ defined as the probability that SINR for both DL and UL exceed predefined thresholds $\tau_\text{DL}$ and $\tau_\text{UL}$, respectively, is given as:
\begin{equation}
\begin{aligned}[]
    J_k &=  \mathbb{E}_{z\in\{{L,N}\}}[\mathbb{P}[\text{SINR}_z^{\text{DL}}>\tau_\text{DL},\text{SINR}_z^{\text{UL}}>\tau_\text{UL} | z, k\in S_t ]]\\
    &= P_L(r_k|k\in S_t)\times J_k^{L} + P_N(r_k|k\in S_t)\times J_k^{N},
    \label{UkFormula}
\end{aligned}
\end{equation}
where
\begin{equation}
\begin{aligned}[]
    J_k^{z} &=  \sum_{j =1}^{m_{z}} \binom{m_z}{j} (-1)^{j +1}\\
     &\exp\left({\frac{-j \eta_z n_0^2 \tau_{\text{DL}}}{P_u G_0 (r_k^2 + h^2)^{-\frac{\alpha_z}{2}}}}\right) \mathcal{L}_{\text{DL}}\left({\frac{-j \eta_z \tau_\text{DL}}{P_u G_0 (r_k^2 + h^2)^{-\frac{\alpha_z}{2}}}}\right)\\
     &\times \sum_{j =1}^{m_{z}} \binom{m_z}{j} (-1)^{j +1}\\
     &\exp\left({\frac{-j \eta_z n_0^2 \tau_{\text{UL}}}{P_d G_0 (r_k^2 + h^2)^{-\frac{\alpha_z}{2}}}}\right) \mathcal{L}_{\text{UL}}\left({\frac{-j \eta_z \tau_\text{UL}}{P_d G_0 (r_k^2 + h^2)^{-\frac{\alpha_z}{2}}}}\right),
    \label{UkFormula2}
\end{aligned}
\end{equation}
where $z\in\{{L, N}\}$, $k\in S_t$ indicates that device $k$ is scheduled for a RB at time $t$, $m_z$ is the Nakagami fading parameter for $z$ link with $\eta_z = m_z(m_z!)^{-1/m_z}$, $\tau_\text{DL}$ and $\tau_\text{UL}$ are SINR thresholds for DL and UL transmissions respectively, and $\mathcal{L}_{\text{DL}}(\cdot)$ and $\mathcal{L}_{\text{UL}}(\cdot)$ are the Laplace transforms of DL and UL inteferences, respectively, and are provided in the next two lemmas.
\begin{IEEEproof}
See Appendix A.
\end{IEEEproof}
\end{theorem}
The Laplace transforms of DL and inter-cluster UL interferences are formulated in the following lemmas:
\begin{lemma}\normalfont
In a UAV-assisted wireless network modeled as an MCP, the Laplace transform of the DL interference is given by:
\begin{equation*}
\begin{aligned}
&\mathcal{L}_{\text{DL}}(s) = \prod_{z\in\{{L, N}\}} \exp\left(-2\pi \lambda \int_0^{\infty} \left[1 - \sum^4_{i=1}p_i\right.\right. \\
&\left.\left.\left(1+\frac{s P_u G_i (g^2+h^2)^{-\frac{\alpha_z}{2}}}{m_z} \right)^{-m_z}\right]qP_z(q)\,dq\right),
\end{aligned}
\end{equation*}
\begin{IEEEproof} 
See Appendix B.
\end{IEEEproof}
\end{lemma}

\begin{lemma}\normalfont
The Laplace transform of the inter-cluster UL interference in a UAV-assisted network modeled as MCP is given by:
\begin{equation*}
\begin{aligned}
\mathcal{L}_{\text{UL}}(s) = \prod_{z\in\{{L, N}\}} \exp&\left(-2\pi \lambda \left[ \int_0^{R} \left[1-[O^z_{e1}(s,q)]^{\Bar{N}} \right]q\,dq\right.\right.\\
&\left.\left. + \int_R^{\infty} \left[1-[O_{e2}^z(s,q)]^{\Bar{N}} \right]q\,dq \vphantom{\int_0^{R} \left[1-[O_{e1}(s,q)]^{\Bar{N}} \right]q\,dq}\right]\right),
\end{aligned}
\end{equation*}
where $\Bar{N}$ represents the number of interfering devices in a neighboring cluster, which is $\Bar{N}=1$ due to the considered scheduling scheme and
\begin{equation*}
\begin{aligned}
O_{e1}^z(s,q) = \int_{|q-R|}^{R+q}\sum^4_{i=1}p_i\left(1+\frac{s P_d G_i (g^2+h^2)^{-\frac{\alpha_z}{2}}}{m_z} \right)^{-m_z} \\
\times \frac{2g}{\pi R^2}\arccos{\left(\frac{g^2+q^2-R^2}{2gq}\right)}\,dq\\
+\int_{0}^{R-q}\sum^4_{i=1}p_i\left(1+\frac{s P_d G_i (g^2+h^2)^{-\frac{\alpha_z}{2}}}{m_z} \right)^{-m_z}\times \frac{2g}{R^2}\,dq,
\end{aligned}
\end{equation*}
and
\begin{equation*}
\begin{aligned}
O_{e2}^z(s,q) = \int_{|q-R|}^{R+q} \sum^4_{i=1}p_i \left(1+\frac{s P_d G_i (g^2+h^2)^{-\frac{\alpha_z}{2}}}{m_z} \right)^{-m_z} \\
\times \frac{2g}{\pi R^2}\arccos{\left(\frac{g^2+q^2-R^2}{2gq}\right)}\,dq.\\
\end{aligned}
\end{equation*}
\begin{IEEEproof} 
See Appendix C.
\end{IEEEproof}
\end{lemma}
\section{Numerical Results}
\begin{table}[]
\centering
\caption{System parameter values}
\label{tab:my-table}
\resizebox{0.95\columnwidth}{!}{%
\begin{tabular}{|l|l|l|}
\hline
$N, M =$ 100, 90 & $R, h=$ 100, 120 m & $\lambda=2/(\pi150^2)~\text{UAV}/\text{m}^2$ \\ \hline
$P_d, P_u=$ 0.1, 0.25 W & $\alpha_L, \alpha_N$ = 2.1, 3.6 & $n_0^2= 4.14\times10^{-6}$ W \\ \hline
$a, b =$ 9.61, 0.16 & $m_L, m_N$ = 3, 1 & $\tau_\text{DL}, \tau_\text{UL} =$ 15, 0 dB \\ \hline
$M_u =$ 10 dB & $M_d=$ 5 dB & $m_u, m_d=$ -1, -3 dB \\ \hline
\end{tabular}%
}
\end{table}

\begin{figure}[t]
    \centering
    \resizebox{0.8\linewidth}{!}{\begin{tikzpicture}[thick,scale=1, every node/.style={scale=1.3},font=\Huge]
        \input{Figures/F2_new2.tex}
        \end{tikzpicture}}
    \captionsetup{font=footnotesize}
    \setlength{\belowcaptionskip}{-10pt}
    \caption{Coverage probability plot for parameter values in Table~\ref{tab:my-table}.}
    \label{fig:success_prob}
\end{figure}
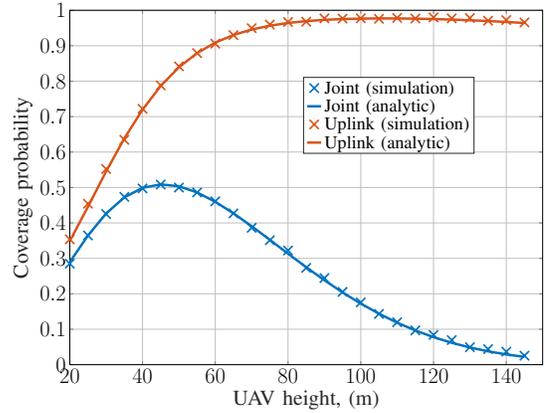

In this section, we provide analytical results and Monte-Carlo simulations to validate the accuracy of the joint success probability given in Theorem 1. Furthermore, we present simulation results for the proposed FL algorithm in UAV-assisted networks. Unless otherwise mentioned, the simulation parameters are presented in Table~\ref{tab:my-table}. We compare our FL aggregator (i.e., that accounts for the joint DL/UL unreliability) with the traditional FedAvg algorithm and with~\cite{salehi2021federated} that considers the UL unreliability only. Fig.~\ref{fig:success_prob} plots the coverage probabilities for both joint DL and UL, and UL-only instances as a function of the UAV height. It can be seen clearly that the results of the analytic expression match the Monte-Carlo simulations, which proves the validity of the conducted analysis. The figure also shows the prominent impact of the DL on the transmission success probability when compared with the UL-only case. Interestingly, the figure shows an optimal height for the UAV, which strikes a tradeoff between improving LOS probability and aggravating propagation loss as height increases.
\begin{figure*}[t]
            \renewcommand{\thefigure}{4a}
            \begin{minipage}[t]{0.33\textwidth}
            \resizebox{0.97\columnwidth}{!}{\begin{tikzpicture}[thick,scale=1, every node/.style={scale=1.3},font=\Huge]
            \input{Figures/F4_loss.tex}
            \end{tikzpicture}}
            \setlength{\belowcaptionskip}{-10pt}
            \captionsetup{font=footnotesize}
            \caption{Global loss across communication rounds.}
            \label{fig:Loss}
            \end{minipage}%
            \renewcommand{\thefigure}{4b}
            \begin{minipage}[t]{0.33\textwidth}
            \resizebox{0.97\columnwidth}{!}{\begin{tikzpicture}[thick,scale=1, every node/.style={scale=1.3},font=\Huge]
            \input{Figures/F4.tex}
            \end{tikzpicture}}
            \setlength{\belowcaptionskip}{-10pt}
            \captionsetup{font=footnotesize}
            \caption{Training accuracy across communication rounds.}
		\label{fig:accuracy}
            \end{minipage}%
            \renewcommand{\thefigure}{5}
            \begin{minipage}[t]{0.33\textwidth}
            \resizebox{0.97\columnwidth}{!}{\begin{tikzpicture}[thick,scale=1, every node/.style={scale=1.3},font=\Huge]
            \input{Figures/DiffE.tex}
            \end{tikzpicture}}
            \setlength{\belowcaptionskip}{-10pt}
            \captionsetup{font=footnotesize}
            \caption{Testing accuracy for different local SGD iteration values.}
		\label{fig:DiffE}
            \end{minipage}%
\end{figure*}

\begin{figure*}[t]
            \renewcommand{\thefigure}{6}
		\begin{minipage}[t]{0.33\textwidth}
			\resizebox{0.97\columnwidth}{!}{\begin{tikzpicture}[thick,scale=1, every node/.style={scale=1.3},font=\Huge]
				\input{Figures/F3_new.tex}
				\end{tikzpicture}}
            \setlength{\belowcaptionskip}{-10pt}
                \captionsetup{font=footnotesize}
			\caption{Training accuracy for different UAV heights.}
			\label{fig:diffHeight}
		\end{minipage}%
             \renewcommand{\thefigure}{7a}
		\begin{minipage}[t]{0.33\textwidth}
			\resizebox{0.97\columnwidth}{!}{\begin{tikzpicture}[thick,scale=1, every node/.style={scale=1.3},font=\Huge]
				\input{Figures/F11_new.tex}
				\end{tikzpicture}}
            \setlength{\belowcaptionskip}{-10pt}
            \captionsetup{font=footnotesize}
            \caption{Training accuracy for different environments at $h=25$ m.}
		\label{fig:Env1}
		\end{minipage}%
             \renewcommand{\thefigure}{7b}
		\begin{minipage}[t]{0.33\textwidth}
			\resizebox{0.97\columnwidth}{!}{\begin{tikzpicture}[thick,scale=1, every node/.style={scale=1.3},font=\Huge]
				\input{Figures/F10_new.tex}
				\end{tikzpicture}}
            \setlength{\belowcaptionskip}{-10pt}
            \captionsetup{font=footnotesize}
            \caption{Training accuracy for different environments at $h=120$ m.}
            \label{fig:Env2}
		\end{minipage}%
\end{figure*}

To evaluate the performance of the proposed FL algorithm, we rely on the global loss and the training and testing accuracies of the global model. We perform FL with the number of local SGD iterations of $E=2$ and batch size of $\xi_t^k = 64$ until the convergence of the global model. Data from the MNIST dataset is distributed in a non-iid manner across all devices within a cluster. 
The result in Fig.~\ref{fig:Loss} shows the global loss versus the number of communication rounds between a UAV and the associated devices. It can be noticed that the proposed method drastically reduces the loss compared to the method based on only erroneous UL channel and the conventional FedAvg. The superiority of the proposed method is also reflected in accuracy plots in Fig.~\ref{fig:accuracy}. The proposed method reaches the convergence much faster and provides around 25\% improvement as compared to the UL-only scenario. The results evidence the importance of accounting for the impact of wireless channels and considering both DL and UL on FL. 

The effect of changing the number of local iterations is studied and the results are given in the bar plot in Fig.~\ref{fig:DiffE}. One can observe that there is an optimal value for local iterations which is $E=2$. At this value, both methods using joint and UL-only success probabilities demonstrate the highest testing accuracies. It is noted that the proposed joint method always performs better than the method relying on only UL success probability.

The developed model allows to investigate the impact of different system parameters on the learning performance. Fig.~\ref{fig:diffHeight} illustrates that the highest accuracy corresponds to the optimal UAV height of $h=50$ m, which is supported by the observations from Fig.~\ref{fig:success_prob}. The reason for such trend is that the higher coverage probability value, the more devices participate in each round of FL.
In Fig.~\ref{fig:Env1} and Fig.~\ref{fig:Env2}, we plot the training accuracy values in four environments for the UAV heights of $h=25$ m and $h=120$ m, respectively. We can notice that the impact of the UAV height on the accuracy of the FL algorithm is not the same for the considered environments. For instance, the FL training accuracy in the suburban environment is the highest for $h=25$ m and lowest for $h=120$ m. The impact of the height in the suburban environment is always negative since the LOS probability is almost 1 even at low heights. Thus, the increase of the height only affects the path-loss that leads to the drop in FL accuracy. We observe an opposite trend for the high-rise urban setting due to a low value of LOS probability at low and a high value at high altitudes. For the high rise urban case, the LOS probability is low at low UAV altitudes, which leads to low FL accuracy. When the UAV flies at higher altitudes, the LOS probability starts to increase. This increase compensates the deterioration caused by increasing path-loss. Thus, the FL accuracy improves for higher UAV heights. Such behavior happens up to some extend after which the path-loss factor dominates again when the LOS probability tends to 1, which results in an accuracy drop. These results signify the need for optimizing the UAV position in addition to other design parameters to improve the FL accuracy in aerial networks, subjects that are expected to be at the forefront of future networks design and analysis.

\vspace{-0.2cm}
\section{Conclusion}
In this paper, we examine the performance of the FL algorithm in aerial UAV-assisted networks. The unreliable and resource-constrained wireless channel between aggregating UAVs and devices makes the learning inefficient due to biasing the global model towards devices with better channel conditions. Using tools from stochastic geometry, the joint upload and download success probability for FL in UAV-assisted networks is computed and used to unbias the FL aggregation rule. The proposed FL algorithm not only surpasses the existing methods but also enables the adjustment of system parameters for optimum learning performance.

\section*{Appendix A}
We derive the joint success probability based on the fact that probabilities of SINR for DL and UL exceeding the thresholds become independent when we condition on the serving distance. The serving distance remains the same for both DL and UL transmissions since we consider a single-tier network where devices always associate with UAVs in the centers of clusters. Thus, after considering LOS and NLOS possibilities for the serving links, the overall joint success probability is derived as
\begin{equation}\small
\begin{aligned}[]
    J_k &=  \mathbb{E}_{z\in\{{L,N}\}}[\mathbb{P}[\text{SINR}_z^{\text{DL}}>\tau_\text{DL},\text{SINR}_z^{\text{UL}}>\tau_\text{UL} | k\in S_t ]]\\
    &= P_L(r_k|k\in S_t)\times J_k^{L} + P_N(r_k|k\in S_t)\times J_k^{N},
\end{aligned}
\end{equation}
where 
\begin{equation}\small
\begin{aligned}[]
    J_k^z &= \mathbb{P}[\text{SINR}_z^\text{DL}>\tau_\text{DL}|k\in S_t]\times \mathbb{P}[\text{SINR}_z^\text{UL}>\tau_\text{UL}|k\in S_t]\\
    &= \mathbb{E}_{I_\text{DL}} \left[\mathbb{P}\left[|h_k|_z^2>{\frac{\tau_{\text{DL}}(n_0^2 + I_\text{DL}) }{P_u G_0 (r_k^2 + h^2)^{-\frac{\alpha_z}{2}}}}\right] \right] \\
    & \ \ \ \ \ \ \ \ \ \ \times \mathbb{E}_{I_\text{UL}} \left[ \mathbb{P}\left[|h_k|_z^2>{\frac{\tau_{\text{UL}}(n_0^2 + I_\text{UL}) }{P_d G_0 (r_k^2 + h^2)^{-\frac{\alpha_z}{2}}}}\right] \right]\\
    &\stackrel{\text{(a)}}{=}\left(1 - \mathbb{E}_{I_{\text{DL}}}\left[\left(1 - \exp\left({\frac{-\eta_z \tau_{\text{DL}}  (n_0^2+I_{\text{DL}})}{P_u G_0 (r_k^2 + h^2)^{-\frac{\alpha_z}{2}}}}\right)\right)^{m_z}\right] \right) \\
    &\times \left(1 - \mathbb{E}_{I_{\text{UL}}}\left[\left(1 - \exp\left({\frac{-\eta_z \tau_{\text{UL}}  (n_0^2+I_{\text{UL}})}{P_d G_0 (r_k^2 + h^2)^{-\frac{\alpha_z}{2}}}}\right)\right)^{m_z}\right] \right) \\
    & \stackrel{\text{(b)}}{=} \sum_{j =1}^{m_{z}} \binom{m_z}{j} (-1)^{j +1}\exp\left({\frac{-j \eta_z n_0^2 \tau_{\text{DL}}}{P_u G_0 (r_k^2 + h^2)^{-\frac{\alpha_z}{2}}}}\right)\\
     &\mathbb{E}_{I_{\text{DL}}}\left[\exp{\left({\frac{-j \eta_z \tau_\text{DL}}{P_u G_0 (r_k^2 + h^2)^{-\frac{\alpha_z}{2}}}}\right)}\right]\\
     &\times \sum_{j =1}^{m_{z}} \binom{m_z}{j} (-1)^{j +1}\exp\left({\frac{-j \eta_z n_0^2 \tau_{\text{UL}}}{P_d G_0 (r_k^2 + h^2)^{-\frac{\alpha_z}{2}}}}\right)\\
     &\mathbb{E}_{I_{\text{UL}}}\left[\exp{\left({\frac{-j \eta_z \tau_\text{UL}}{P_d G_0 (r_k^2 + h^2)^{-\frac{\alpha_z}{2}}}}\right)}\right],
\end{aligned}
\end{equation}

where (a) is the result of using the cumulative distribution function (CDF) of the normalized gamma fading coefficient, (b) is obtained by using the binomial expansion. The final result is obtained by defining the expectations as the Laplace transforms of DL and UL interference.

\section*{Appendix B}
The overall DL intereference consists of LOS and NLOS components: $I_{\text{DL}}=I_{\text{DL}}^L + I_{\text{DL}}^N$. When evaluating the exponential function for $I_{\text{DL}}$, the result becomes the product of the two components. Thus, the Laplace transform of the DL interference can be derived as
\begin{equation*}\small
\begin{aligned}
&\mathcal{L}_{\text{DL}}(s) \\
&=\prod_{z\in\{{L, N}\}} \mathbb{E}\left[\exp{\left(-s\sum_{u\in \Phi^z\backslash u_0}P_u G_u |h_u|_z^2 (q^2+h^2)^{-\frac{\alpha_z}{2}} \right)} \right]\\
&= \prod_{z\in\{{L, N}\}} \mathbb{E}_{\Phi^z}\left[\prod_{u\in \Phi^z\backslash u_0}  \mathbb{E}_{\substack{G_u,\\|h_u|_z^2}}\left[\exp\left(-s P_u G_u |h_u|_z^2 \right. \right.\right. \\
&\ \ \ \ \ \ \ \ \ \ \ \ \ \ \ \ \ \ \ \ \ \ \ \ \ \ \ \ \ \ \ \ \left. \left. \left. \times(q^2+h^2)^{-\frac{\alpha_z}{2}}\right) \right]\right]\\
&\stackrel{\text{(a)}}{=}\prod_{z\in\{{L, N}\}} \exp\left(-2\pi \lambda \int_0^{\infty} \left[1 - \mathbb{E}_{\substack{G_u,\\|h_u|_z^2}}\left[\exp\left(-s P_u G_u |h_u|_z^2 \right.\right.\right.\right.\\
&\ \ \ \ \ \ \ \ \ \ \ \ \ \ \ \ \ \ \ \ \ \ \ \ \ \ \ \ \ \ \ \ \left.\left. \left.\left.  \times (q^2+h^2)^{-\frac{\alpha_z}{2}}\right) \right]\right]qP_z(q)\,dq\right),
\end{aligned}
\end{equation*}
where (a) is the result of applying the probability generating functional of PPP for UAV locations. The final form is obtained from applying the moment generating function of the gamma distribution characterizing the small-scale fading gain and from averaging over the dimensionality gain using the probability mass function (PMF) in Table~\ref{tab:my-table2}.

\section*{Appendix C}
Similar to DL interference, the UL intereference is comprised of two compoenents: $I_{\text{UL}}=I_{\text{UL}}^L + I_{\text{UL}}^N$. Thus, the Laplace transform of the UL interference can be derived as
\begin{equation*}\small
\begin{aligned}
&\mathcal{L}_{\text{UL}}(s) =\prod_{z\in\{{L, N}\}} \mathbb{E}\left[\exp\left(-s\sum_{u\in \Phi^z\backslash u_0} \sum_{d\in D_u} P_d G_d |h_d|_z^2 \right. \right.\\
& \ \ \ \ \ \ \ \ \ \ \ \ \ \ \ \ \ \ \ \ \ \ \ \ \ \ \ \ \ \ \ \ \ \left.\left. \times (g^2+h^2)^{-\frac{\alpha_z}{2}} \right) \right] \\
&\stackrel{\text{(a)}}{=} \prod_{z\in\{{L, N}\}} \mathbb{E}_{u}\left[\prod_{u\in \Phi^z\backslash u_0} \mathbb{E}_{d}\left[ \prod_{d\in D_u}\sum^4_{i=1}p_i  \left(1 \right. \right.  \right. \\
& \ \ \ \ \ \ \ \ \ \ \ \ \ \ \ \ \ \ \ \ \ \ \ \ \ \ \ \ \left.\left. \left.+\frac{s P_d G_i (g^2+h^2)^{-\frac{\alpha_z}{2}}}{m_z} \right)^{-m_z} \middle|q\right]\right]\\
&\stackrel{\text{(b)}}{=}\prod_{z\in\{{L, N}\}}\exp{\left(-2\pi \lambda_v \int_0^{\infty} \left[1-[O_e^z(s,q)]^{\Bar{N}} \right]qP_z(q)\,dq \right)}\\
\end{aligned}
\end{equation*}
where
\begin{equation*}
\begin{aligned}
 O_{e}^z(s,q) &= \int_0^{\infty}\sum^4_{i=1}p_i \left(1+\frac{s P_d G_i (g^2+h^2)^{-\frac{\alpha_z}{2}}}{m_z} \right)^{-m_z} \\
 &\ \ \ \ \ \ \ \ \ \ \ \ \ \ \ \  \ \ \ \ \ \ \ \ \ \ \ \ \ \ \times f_Q^{\text{MAT}}(g|q)\,dg\vphantom{\left(1+\frac{s P_d (g^2+h^2)^{-\frac{\alpha}{2}}}{m} \right)^{-m}},
\end{aligned}
\end{equation*}
where (a) is obtained by applying the moment generating function of the small-scale fading gain. (b) is the result of the probability generating functional of PPP for UAV locations. The final result follows from applying the piece-wise expression of the conditional distance distribution $f_Q^{\text{MAT}}(g|q)$ given in (\ref{condDist}).

\bibliographystyle{IEEEtran}
\bibliography{reference}
\end{document}

%% file: Figures/defs_tikzpgf.tex
\usepackage{tikz}
\usepackage{pgfplotstable}
\usepackage{rotate}

\definecolor{rubblue}{cmyk}{1,0.5,0,0.6}
\definecolor{rubgreen}{cmyk}{0.5,0,1,0}
\definecolor{rubgray}{cmyk}{0.03,0.03,0.03,0.1}

\usepgfplotslibrary{units}

\usetikzlibrary{%
patterns,%
calc,%
fit,%
arrows,%
plotmarks,%
shadows,%
chains,%
shapes%
}

\tikzset{>=latex'} 
\tikzstyle{every picture}+=[remember picture] 
\pgfdeclarelayer{background}
\pgfdeclarelayer{foreground}
\pgfsetlayers{background,main,foreground}

\tikzstyle{blueblock}=[draw=rubblue, rectangle, thick, drop shadow, minimum width=20mm, minimum height=8mm,fill=rubblue!20, text width=20mm, text centered]
\tikzstyle{bluebox}=[draw=rubblue, rectangle, thick, drop shadow, minimum width=8mm, minimum height=8mm,fill=rubblue!20, text width=8mm, text centered]
\tikzstyle{greenblock}=[draw=rubgreen, rectangle, thick, drop shadow, minimum width=20mm, minimum height=8mm,fill=rubgreen!20, text width=20mm, text centered]
\tikzstyle{dot} = [draw, circle, minimum size=0.2pt,scale=0.3,fill=black,black]
\tikzstyle{smalldot} = [draw, circle, minimum size=0.1pt,scale=0.2,fill=black,black]
\tikzstyle{reddot}  =[draw,circle,minimum size=0.2pt,scale=0.8,fill=red,thin]
\tikzstyle{greendot}  =[draw,circle,minimum size=0.2pt,scale=0.8,fill=Green,thin]
\tikzstyle{bluedot}  =[draw,circle,minimum size=0.2pt,scale=0.8,fill=blue,thin]
\tikzstyle{whitedot}=[draw,circle,minimum size=0.2pt,scale=0.8,fill=white,thin]
\tikzstyle{blackdot} = [draw, circle, minimum size=0.2pt,scale=0.7,fill=black,black]
\tikzstyle{sum} = [drop shadow, draw=rubblue, thick, fill=rubblue!20, circle]
\tikzstyle{relay} = [blueblock, minimum width=5mm, minimum height=20mm, text width=5mm, rounded corners=2pt]
\tikzstyle{relay2} = [blueblock, minimum width=5mm, minimum height=15mm, text width=5mm, rounded corners=2pt]
\tikzstyle{relay3} = [blueblock, minimum width=5mm, minimum height=25mm, text width=5mm, rounded corners=2pt]
\tikzstyle{relay4} = [blueblock, minimum width=5mm, minimum height=10mm, text width=5mm, rounded corners=2pt]
\tikzstyle{relay5} = [blueblock, minimum width=5mm, minimum height=50mm, text width=5mm, rounded corners=2pt]
\tikzstyle{relay6} = [blueblock, minimum width=5mm, minimum height=5mm, text width=5mm, rounded corners=2pt]
\tikzstyle{circgreen} = [draw, circle, inner sep=2pt, fill=rubgreen, drop shadow, thick]
\tikzstyle{circwhite} = [draw, circle, inner sep=2pt, fill=white, drop shadow, thick]
\tikzstyle{circdashed} = [draw, dashed, circle, inner sep=2pt, fill=rubgray, drop shadow, thick]
\tikzstyle{vertbox} = [rectangle, draw=rubblue, thick, rotate=90, text centered, minimum width=16.5mm, minimum height=8mm, text width=16.5mm, inner sep=0pt, fill=rubblue!20, drop shadow]
\tikzstyle{vertboxb} = [rectangle, draw=rubblue, thick, rotate=90, text centered, minimum width=16.5mm, minimum height=8mm, text width=16.5mm, fill=rubblue!20, drop shadow]
\tikzstyle{vertboxshort} = [rectangle, draw=rubblue, thick, rotate=90, text centered, minimum width=10mm, minimum height=8mm, text width=10mm, inner sep=0pt, fill=rubblue!20, drop shadow]
\tikzstyle{smalldotgreen} = [draw=rubgreen, circle, minimum size=0.2pt,scale=0.8,fill=rubgreen!20]
\tikzstyle{antenna} = [regular polygon, regular polygon sides=3, draw, shape border rotate=180, minimum size=0.2pt, scale=0.3]

\tikzstyle{poly} = [regular polygon, regular polygon sides=6, shape aspect=0.5, minimum width=1.5cm, minimum height=0.35cm, draw, dashed]

\definecolor{cff9e00}{RGB}{255,158,0}
\definecolor{c4fff00}{RGB}{79,255,0}
\definecolor{cff0012}{RGB}{255,0,18}
\definecolor{c00c5ff}{RGB}{0,197,255}
\definecolor{c046f00}{RGB}{4,111,0}
\definecolor{c004b9d}{RGB}{0,75,157}

\newlength{\mylen}

%% file: Figures/F2_new2.tex
%
%

\definecolor{mycolor1}{rgb}{0.00000,0.44700,0.74100}%
\definecolor{mycolor2}{rgb}{0.85000,0.32500,0.09800}%
\definecolor{mycolor3}{rgb}{0.92900,0.69400,0.12500}%
\definecolor{mycolor4}{rgb}{0.49400,0.18400,0.55600}%
\definecolor{mycolor5}{rgb}{0.74902,0.00000,0.74902}%
\definecolor{mycolor6}{rgb}{0.00000,0.44706,0.74118}%
\definecolor{mycolor7}{rgb}{0.63500,0.07800,0.18400}
\definecolor{mycolor8}{rgb}{0.46667,0.67451,0.18824}%

\begin{axis}[%
width=8in,
height=6in,
at={(0.78in,0.556in)},
scale only axis,
xmin=20,
xmax=150,
ytick={0,0.1,...,1},
xtick={20,40,...,140},
ticklabel style={font=\huge},
xlabel style={font=\color{white!15!black}},
xlabel={\huge{$\text{UAV height, (m)}$}},
ymin=0,
ymax=1,
ylabel style={font=\color{white!15!black}},
ylabel={\huge{$\text{Coverage probability} $}},
axis background/.style={fill=white},
xmajorgrids,
ymajorgrids,
legend style={at={(0.494,0.596)}, anchor=south west, legend cell align=left, align=left, draw=white!15!black,font=\LARGE}
]
\addplot [color=mycolor1, line width=2.0pt,only marks,mark size=8.0pt, mark=x, mark options={solid, mycolor1}]
  table[row sep=crcr]{%
20	0.285\\
25	0.3646\\
30	0.425\\
35	0.4736\\
40	0.4974\\
45	0.5085\\
50	0.4998\\
55	0.486\\
60	0.461\\
65	0.427\\
70	0.3862\\
75	0.3512\\
80	0.3222\\
85	0.2732\\
90	0.2444\\
95	0.2054\\
100	0.1758\\
105	0.1432\\
110	0.1198\\
115	0.0968\\
120	0.084\\
125	0.0694\\
130	0.0494\\
135	0.044\\
140	0.0368\\
145	0.0254\\
};
\addlegendentry{Joint (simulation)}

\addplot [color=mycolor1, line width=3.0pt, mark repeat={20}, mark size=6.0pt, mark options={solid, mycolor1}]
  table[row sep=crcr]{%
20	0.289398247675922\\
25	0.364559872177232\\
30	0.426759472536274\\
35	0.472053341785117\\
40	0.499101701516407\\
45	0.508770089150336\\
50	0.503405158977419\\
55	0.486030146023136\\
60	0.45970946489632\\
65	0.427188719908788\\
70	0.39077064462156\\
75	0.352326749712465\\
80	0.313358041949825\\
85	0.27505698642892\\
90	0.238354353469853\\
95	0.203951297472528\\
100	0.172342652987968\\
105	0.143837357884647\\
110	0.118579853612485\\
115	0.0965740168797123\\
120	0.0777094256098275\\
125	0.0617887109357091\\
130	0.0485543467218703\\
135	0.0377132978829179\\
140	0.0289583115602478\\
145	0.0219851239701783\\
};
\addlegendentry{Joint (analytic)}

\addplot [color=mycolor2, line width=2.0pt,only marks,mark size=8.0pt, mark=x, mark options={solid, mycolor2}]
  table[row sep=crcr]{%
20	0.3536\\
25	0.4542\\
30	0.5522\\
35	0.6346\\
40	0.722\\
45	0.7882\\
50	0.842\\
55	0.8792\\
60	0.9058\\
65	0.93\\
70	0.9496\\
75	0.9598\\
80	0.967\\
85	0.9678\\
90	0.9772\\
95	0.9766\\
100	0.9774\\
105	0.976\\
110	0.9786\\
115	0.9768\\
120	0.98\\
125	0.9772\\
130	0.9782\\
135	0.9706\\
140	0.9724\\
145	0.9658\\
};
\addlegendentry{Uplink (simulation)}

\addplot [color=mycolor2, line width=3.0pt, mark repeat={20}, mark size=6.0pt, mark options={solid, mycolor2}]
  table[row sep=crcr]{%
20	0.348832050861429\\
25	0.4457842879812\\
30	0.545436887798604\\
35	0.638562906590123\\
40	0.71972017776423\\
45	0.786612132244895\\
50	0.839288151096023\\
55	0.879295679260452\\
60	0.908870534763558\\
65	0.930320950827173\\
70	0.945676625641853\\
75	0.956560856793397\\
80	0.964197248083197\\
85	0.969474673028557\\
90	0.973024446102457\\
95	0.975288953176677\\
100	0.976575659929765\\
105	0.977096989914615\\
110	0.976998753054319\\
115	0.976380051964653\\
120	0.975307125678412\\
125	0.973822952852\\
130	0.971953949115951\\
135	0.969714667745859\\
140	0.967111141464315\\
145	0.964143286030566\\
};
\addlegendentry{Uplink (analytic)}

\end{axis}

\begin{axis}[%
width=6in,
height=4.8in,
at={(0in,0in)},
scale only axis,
xmin=0,
xmax=1,
ymin=0,
ymax=1,
axis line style={draw=none},
ticks=none,
axis x line*=bottom,
axis y line*=left
]
\end{axis}

%% file: Figures/F4_loss.tex
%
%
\definecolor{mycolor1}{rgb}{0.00000,0.44700,0.74100}%
\definecolor{mycolor2}{rgb}{0.85000,0.32500,0.09800}%
\definecolor{mycolor3}{rgb}{0.92900,0.69400,0.12500}%
\definecolor{mycolor4}{rgb}{0.49400,0.18400,0.55600}%
\definecolor{mycolor5}{rgb}{0.74902,0.00000,0.74902}%
\definecolor{mycolor6}{rgb}{0.00000,0.44706,0.74118}%
\definecolor{mycolor7}{rgb}{0.63500,0.07800,0.18400}

\begin{axis}[%
width=8in,
height=7in,
at={(0.78in,0.556in)},
scale only axis,
xmin=0,
xmax=255,
xtick={0,50,...,250},
ytick={0.6,0.8,...,2.2},
xlabel style={font=\color{white!15!black}},
xlabel={\Huge{$\text{Communication rounds}$}},
ymin=0.594741454864154,
ymax=2.39474145486415,
ylabel style={font=\color{white!15!black}},
ylabel={\Huge{$\text{Global loss} $}},
axis background/.style={fill=white},
xmajorgrids,
ymajorgrids,
legend style={at={(0.605,0.468)}, anchor=south west, legend cell align=left, align=left, draw=white!15!black,font=\huge}
]
\addplot [color=mycolor1, line width=3.0pt, mark repeat={20}, mark size=6.0pt, mark=o, mark options={solid, mycolor1}]
  table[row sep=crcr]{%
1	2.33837955474854\\
2	2.312\\
3	2.30689725875855\\
4	2.15759453773499\\
5	2.08649518489838\\
6	2.01534728050232\\
7	1.94146120071411\\
8	1.8814675951004\\
9	1.81178284168243\\
10	1.75974709033966\\
11	1.70991616725922\\
12	1.657913646698\\
13	1.60704517364502\\
14	1.56585300922394\\
15	1.51995086669922\\
16	1.49033908367157\\
17	1.45886109828949\\
18	1.42519796848297\\
19	1.39596122741699\\
20	1.37381548404694\\
21	1.34613340854645\\
22	1.32652007579803\\
23	1.3008491230011\\
24	1.28731580257416\\
25	1.27875063419342\\
26	1.25465930938721\\
27	1.23124344825745\\
28	1.21222987174988\\
29	1.19936567306519\\
30	1.18264260292053\\
31	1.17380686998367\\
32	1.16357573032379\\
33	1.14952172279358\\
34	1.14433868169785\\
35	1.12197997093201\\
36	1.10618041038513\\
37	1.09944165468216\\
38	1.08986441612244\\
39	1.08343581199646\\
40	1.06852748632431\\
41	1.05779190778732\\
42	1.05512033224106\\
43	1.0426570057869\\
44	1.03238028764725\\
45	1.02271148204803\\
46	1.01696335077286\\
47	1.01168635129929\\
48	1.0064945101738\\
49	0.995750634670258\\
50	0.992444367408752\\
51	0.985876705646515\\
52	0.981009070873261\\
53	0.973819296360016\\
54	0.968342487812042\\
55	0.962246475219727\\
56	0.95476361989975\\
57	0.948280119895935\\
58	0.943939335346222\\
59	0.938418736457825\\
60	0.934288375377655\\
61	0.929109110832214\\
62	0.926205468177795\\
63	0.921674253940582\\
64	0.91969934463501\\
65	0.914768424034119\\
66	0.909593632221222\\
67	0.90545120716095\\
68	0.90104953289032\\
69	0.895896844863892\\
70	0.891377444267273\\
71	0.887274281978607\\
72	0.886890099048614\\
73	0.883601877689362\\
74	0.881149446964264\\
75	0.877791850566864\\
76	0.871143186092377\\
77	0.866565616130829\\
78	0.863753283023834\\
79	0.861320333480835\\
80	0.857956418991089\\
81	0.855402932167053\\
82	0.85362101316452\\
83	0.848206522464752\\
84	0.845199475288391\\
85	0.844430961608887\\
86	0.839255511760712\\
87	0.836698620319366\\
88	0.837043218612671\\
89	0.833992795944214\\
90	0.831053748130798\\
91	0.828122715950012\\
92	0.824382829666138\\
93	0.821864128112793\\
94	0.81954859495163\\
95	0.818801620006561\\
96	0.816045637130737\\
97	0.813938393592834\\
98	0.812378449440002\\
99	0.809186379909515\\
100	0.808310880661011\\
101	0.805044822692871\\
102	0.804690403938293\\
103	0.80262512922287\\
104	0.799911680221558\\
105	0.796455609798431\\
106	0.795690100193024\\
107	0.794399583339691\\
108	0.792582468986511\\
109	0.789842352867126\\
110	0.787537868022919\\
111	0.78701961517334\\
112	0.784836056232452\\
113	0.783804771900177\\
114	0.781870536804199\\
115	0.780109677314758\\
116	0.77887717962265\\
117	0.778267874717712\\
118	0.775685091018677\\
119	0.773107485771179\\
120	0.771281287670135\\
121	0.76964498758316\\
122	0.768435423374176\\
123	0.767877695560455\\
124	0.765009984970093\\
125	0.764300987720489\\
126	0.762960638999939\\
127	0.760639967918396\\
128	0.758931212425232\\
129	0.756418895721436\\
130	0.755012354850769\\
131	0.753098840713501\\
132	0.751415934562683\\
133	0.750908324718475\\
134	0.749802045822144\\
135	0.748967249393463\\
136	0.747963502407074\\
137	0.747357995510101\\
138	0.746233470439911\\
139	0.744519956111908\\
140	0.743029744625092\\
141	0.74271942615509\\
142	0.741144905090332\\
143	0.739788682460785\\
144	0.738949038982391\\
145	0.738005571365356\\
146	0.737566885948181\\
147	0.737081933021545\\
148	0.735967118740082\\
149	0.734831576347351\\
150	0.732785477638245\\
151	0.731181180477142\\
152	0.729756324291229\\
153	0.728615090847015\\
154	0.727254877090454\\
155	0.726601366996765\\
156	0.726021218299866\\
157	0.725002119541168\\
158	0.724303488731384\\
159	0.722859408855438\\
160	0.721393101215363\\
161	0.721087398529053\\
162	0.720431087017059\\
163	0.719949486255646\\
164	0.718764681816101\\
165	0.717420163154602\\
166	0.715910477638245\\
167	0.716001455783844\\
168	0.715125722885132\\
169	0.713527770042419\\
170	0.712958474159241\\
171	0.711689240932465\\
172	0.711317012310028\\
173	0.710227460861206\\
174	0.70875143289566\\
175	0.707371995449066\\
176	0.70689474105835\\
177	0.706925506591797\\
178	0.705751147270203\\
179	0.704160203933716\\
180	0.703348217010498\\
181	0.703042652606964\\
182	0.702217762470245\\
183	0.701553537845612\\
184	0.700712208747864\\
185	0.700395274162293\\
186	0.699367365837097\\
187	0.698602840900421\\
188	0.697828884124756\\
189	0.696792633533478\\
190	0.696338336467743\\
191	0.695782735347748\\
192	0.695681746006012\\
193	0.695008707046509\\
194	0.694042413234711\\
195	0.692628569602966\\
196	0.691871809959412\\
197	0.691017792224884\\
198	0.690216493606567\\
199	0.68925369977951\\
200	0.688410036563873\\
201	0.687776129245758\\
202	0.686877608299255\\
203	0.686697261333466\\
204	0.685888590812683\\
205	0.685148313045502\\
206	0.685057315826416\\
207	0.684707164764404\\
208	0.684236328601837\\
209	0.683183660507202\\
210	0.682778573036194\\
211	0.681410927772522\\
212	0.680215466022491\\
213	0.680071461200714\\
214	0.679649419784546\\
215	0.679154667854309\\
216	0.678005032539368\\
217	0.677736527919769\\
218	0.6773362159729\\
219	0.676479332447052\\
220	0.675783092975616\\
221	0.674823482036591\\
222	0.674475719928741\\
223	0.673576672077179\\
224	0.673012602329254\\
225	0.672642076015472\\
226	0.672274444103241\\
227	0.671778030395508\\
228	0.671033823490143\\
229	0.670443215370178\\
230	0.669905030727386\\
231	0.669364564418793\\
232	0.668879907131195\\
233	0.668341884613037\\
234	0.66797568321228\\
235	0.667176313400269\\
236	0.666464123725891\\
237	0.666095414161682\\
238	0.665521290302277\\
239	0.665192134380341\\
240	0.664501292705536\\
241	0.663620300292969\\
242	0.663151760101318\\
243	0.662265231609344\\
244	0.661849236488342\\
245	0.661607356071472\\
246	0.661326274871826\\
247	0.660753834247589\\
248	0.660293118953705\\
249	0.659593269824982\\
250	0.659204647541046\\
251	0.658612191677094\\
};
\addlegendentry{Joint}

\addplot [color=mycolor2, line width=3.0pt, mark repeat={20}, mark size=6.0pt, mark=square, mark options={solid, mycolor2}]
  table[row sep=crcr]{%
1	2.33947243690491\\
2	2.33321251869202\\
3	2.29990119934082\\
4	2.27404246330261\\
5	2.25271732807159\\
6	2.24093430042267\\
7	2.22888994216919\\
8	2.21689734458923\\
9	2.20454258918762\\
10	2.19529824256897\\
11	2.18667271137238\\
12	2.17754912376404\\
13	2.16878833770752\\
14	2.16193115711212\\
15	2.15488426685333\\
16	2.14966428279877\\
17	2.14471924304962\\
18	2.13825767040253\\
19	2.1319310426712\\
20	2.12683367729187\\
21	2.12036616802216\\
22	2.11513366699219\\
23	2.11063063144684\\
24	2.10657844543457\\
25	2.10274567604065\\
26	2.0987893819809\\
27	2.09492166042328\\
28	2.090691614151\\
29	2.08741269111633\\
30	2.08378946781158\\
31	2.07993233203888\\
32	2.0767514705658\\
33	2.07268943786621\\
34	2.06964869499207\\
35	2.06582601070404\\
36	2.06331281661987\\
37	2.06060965061188\\
38	2.05729758739471\\
39	2.05449452400208\\
40	2.05185134410858\\
41	2.04869253635407\\
42	2.04571385383606\\
43	2.04358403682709\\
44	2.04068329334259\\
45	2.03825409412384\\
46	2.03621333837509\\
47	2.03403784036636\\
48	2.03157807588577\\
49	2.0291793346405\\
50	2.02714096307754\\
51	2.0246374130249\\
52	2.02222393751144\\
53	2.02004151344299\\
54	2.0180321931839\\
55	2.01575813293457\\
56	2.01380087137222\\
57	2.01190695762634\\
58	2.01018096208572\\
59	2.00874298810959\\
60	2.00675648450851\\
61	2.00502341985703\\
62	2.00279998779297\\
63	2.00100207328796\\
64	1.99916954040527\\
65	1.99735585451126\\
66	1.99589080810547\\
67	1.99419029951096\\
68	1.99269461631775\\
69	1.991346347332\\
70	1.99007059335709\\
71	1.98846606016159\\
72	1.98690552711487\\
73	1.98501472473145\\
74	1.98337869644165\\
75	1.98165835142136\\
76	1.98004631996155\\
77	1.97864437103271\\
78	1.97733548879623\\
79	1.97602955102921\\
80	1.97474522590637\\
81	1.97337735891342\\
82	1.97184655666351\\
83	1.97050930261612\\
84	1.96915853023529\\
85	1.9677373290062\\
86	1.96637277603149\\
87	1.96502689123154\\
88	1.96361665725708\\
89	1.96239891052246\\
90	1.96109946966171\\
91	1.9599974155426\\
92	1.95862379074097\\
93	1.95754832029343\\
94	1.95654081106186\\
95	1.95543464422226\\
96	1.95435348749161\\
97	1.95329489707947\\
98	1.95204725265503\\
99	1.9508162021637\\
100	1.94963604211807\\
101	1.94852330684662\\
102	1.94736558198929\\
103	1.94638789892197\\
104	1.94536509513855\\
105	1.94431980848312\\
106	1.94336168766022\\
107	1.94239221811295\\
108	1.94134761095047\\
109	1.94036418199539\\
110	1.93924126625061\\
111	1.93824456930161\\
112	1.93712195158005\\
113	1.9362560749054\\
114	1.93516073226929\\
115	1.93414044380188\\
116	1.93329237699509\\
117	1.93225269317627\\
118	1.9311754822731\\
119	1.93038028478622\\
120	1.92932484149933\\
121	1.92846492528915\\
122	1.92765221595764\\
123	1.92669732570648\\
124	1.9259313583374\\
125	1.92503776550293\\
126	1.92396508455276\\
127	1.92294117212296\\
128	1.92210704088211\\
129	1.92133169174194\\
130	1.92053226232529\\
131	1.91960319280624\\
132	1.91883411407471\\
133	1.9179111123085\\
134	1.91710213422775\\
135	1.91623375415802\\
136	1.91552193164825\\
137	1.91470420360565\\
138	1.91375958919525\\
139	1.91299357414246\\
140	1.91215590238571\\
141	1.91136939525604\\
142	1.91055797338486\\
143	1.90967913866043\\
144	1.90902118682861\\
145	1.90818915367126\\
146	1.90750138759613\\
147	1.90680394172668\\
148	1.906003510952\\
149	1.90528637170792\\
150	1.90449995994568\\
151	1.90380125045776\\
152	1.90315873622894\\
153	1.90246931314468\\
154	1.90182328224182\\
155	1.90116757154465\\
156	1.90038796663284\\
157	1.89977744817734\\
158	1.89901111125946\\
159	1.89835766553879\\
160	1.89768482446671\\
161	1.89685143232346\\
162	1.89622001647949\\
163	1.89553005695343\\
164	1.89500452280045\\
165	1.89423977136612\\
166	1.89350589513779\\
167	1.89284331798553\\
168	1.89216281175613\\
169	1.89151841402054\\
170	1.89095579385757\\
171	1.89044089317322\\
172	1.88981157541275\\
173	1.88913758993149\\
174	1.88845820426941\\
175	1.88773931264877\\
176	1.88713014125824\\
177	1.88646261692047\\
178	1.88594115972519\\
179	1.885409283638\\
180	1.88469800949097\\
181	1.88416709899902\\
182	1.88357471227646\\
183	1.88297330141068\\
184	1.88235179185867\\
185	1.88172320127487\\
186	1.88118234872818\\
187	1.88052705526352\\
188	1.88008824586868\\
189	1.87948375940323\\
190	1.87880107164383\\
191	1.87814527750015\\
192	1.87754232883453\\
193	1.8768902182579\\
194	1.87627527713776\\
195	1.87574721574783\\
196	1.87520220279694\\
197	1.8746621966362\\
198	1.87406905889511\\
199	1.87354900836945\\
200	1.87309926748276\\
201	1.87258596420288\\
202	1.8720712184906\\
203	1.87159572839737\\
204	1.87106584310532\\
205	1.87058135271072\\
206	1.8701162815094\\
207	1.86950087547302\\
208	1.86898233890533\\
209	1.86843590736389\\
210	1.86795760393143\\
211	1.86751925945282\\
212	1.86699770689011\\
213	1.86652237176895\\
214	1.8660545706749\\
215	1.86548291444778\\
216	1.86490118503571\\
217	1.86443457603455\\
218	1.8639829158783\\
219	1.86342018842697\\
220	1.86290844678879\\
221	1.86237688064575\\
222	1.86193939447403\\
223	1.86150960922241\\
224	1.8610381603241\\
225	1.86054099798203\\
226	1.86005185842514\\
227	1.85950000286102\\
228	1.85902787446976\\
229	1.8585242152214\\
230	1.85801690816879\\
231	1.85751085281372\\
232	1.85702239274979\\
233	1.85659694671631\\
234	1.85621815919876\\
235	1.85577104091644\\
236	1.85531197786331\\
237	1.85479165315628\\
238	1.85436993837357\\
239	1.85389394760132\\
240	1.85347506999969\\
241	1.85313680171967\\
242	1.85265159606934\\
243	1.85224671363831\\
244	1.85185135602951\\
245	1.85140290260315\\
246	1.85091717243195\\
247	1.85051798820496\\
248	1.85006988048553\\
249	1.849596118927\\
250	1.84920952320099\\
251	1.84873933792114\\
};
\addlegendentry{UL-only}

\addplot [color=mycolor3, line width=3.0pt, mark repeat={20}, mark size=6.0pt, mark=diamond, mark options={solid, mycolor3}]
  table[row sep=crcr]{%
1	2.34204750061035\\
2	2.30202279090881\\
3	2.30250558853149\\
4	2.30261287689209\\
5	2.30259323120117\\
6	2.30260419845581\\
7	2.30259118080139\\
8	2.30257687568665\\
9	2.30259952545166\\
10	2.3025984287262\\
11	2.30257167816162\\
12	2.30257368087769\\
13	2.30259232521057\\
14	2.30259599685669\\
15	2.30259013175964\\
16	2.30259532928467\\
17	2.30258784294128\\
18	2.30259084701538\\
19	2.30258359909058\\
20	2.30257225036621\\
21	2.30259714126587\\
22	2.30258893966675\\
23	2.30258793830872\\
24	2.30259246826172\\
25	2.30258026123047\\
26	2.30258555412292\\
27	2.30258107185364\\
28	2.30258250236511\\
29	2.30257730484009\\
30	2.30258197784424\\
31	2.30258140563965\\
32	2.30257859230041\\
33	2.30258588790894\\
34	2.30258197784424\\
35	2.30258407592773\\
36	2.30258727073669\\
37	2.30258498191834\\
38	2.30258364677429\\
39	2.30258827209473\\
40	2.30257849693298\\
41	2.30258455276489\\
42	2.30258383750916\\
43	2.30258684158325\\
44	2.30258383750916\\
45	2.30258536338806\\
46	2.30258626937866\\
47	2.30258779525757\\
48	2.30258646011353\\
49	2.30258288383484\\
50	2.302583360672\\
51	2.30258474349976\\
52	2.30258350372314\\
53	2.3025857925415\\
54	2.30258231163025\\
55	2.3025842666626\\
56	2.30258622169495\\
57	2.3025851726532\\
58	2.30258288383484\\
59	2.30258498191834\\
60	2.30258512496948\\
61	2.30258574485779\\
62	2.3025821685791\\
63	2.30258498191834\\
64	2.30258297920227\\
65	2.30258512496948\\
66	2.30258564949036\\
67	2.30258650779724\\
68	2.3025848865509\\
69	2.30258321762085\\
70	2.30258264541626\\
71	2.30258274078369\\
72	2.3025860786438\\
73	2.30258588790894\\
74	2.30258646011353\\
75	2.30258417129517\\
76	2.30258493423462\\
77	2.30258636474609\\
78	2.30258584022522\\
79	2.30258531570435\\
80	2.30258688926697\\
81	2.30258588790894\\
82	2.30258469581604\\
83	2.30258493423462\\
84	2.30258541107178\\
85	2.3025860786438\\
86	2.3025851726532\\
87	2.30258727073669\\
88	2.30258646011353\\
89	2.30258417129517\\
90	2.30258331298828\\
91	2.3025857925415\\
92	2.3025851726532\\
93	2.30258383750916\\
94	2.30258507728577\\
95	2.30258512496948\\
96	2.30258474349976\\
97	2.30258526802063\\
98	2.30258741378784\\
99	2.3025839805603\\
100	2.30258316993713\\
101	2.30258560180664\\
102	2.30258455276489\\
103	2.30258636474609\\
104	2.30258502960205\\
105	2.30258283615112\\
106	2.30258502960205\\
107	2.30258445739746\\
108	2.30258555412292\\
109	2.30258588790894\\
110	2.30258512496948\\
111	2.30258469581604\\
112	2.30258498191834\\
113	2.30258555412292\\
114	2.30258421897888\\
115	2.30258502960205\\
116	2.30258555412292\\
117	2.30258674621582\\
118	2.30258483886719\\
119	2.30258564949036\\
120	2.30258498191834\\
121	2.30258545875549\\
122	2.30258512496948\\
123	2.30258364677429\\
124	2.3025860786438\\
125	2.30258388519287\\
126	2.30258388519287\\
127	2.3025860786438\\
128	2.30258512496948\\
129	2.30258564949036\\
130	2.30258450508118\\
131	2.30258502960205\\
132	2.30258655548096\\
133	2.3025851726532\\
134	2.30258531570435\\
135	2.30258479118347\\
136	2.30258479118347\\
137	2.30258455276489\\
138	2.30258555412292\\
139	2.30258388519287\\
140	2.30258407592773\\
141	2.30258703231812\\
142	2.30258545875549\\
143	2.30258421897888\\
144	2.30258507728577\\
145	2.30258588790894\\
146	2.30258612632751\\
147	2.30258598327637\\
148	2.30258421897888\\
149	2.3025839805603\\
150	2.3025848865509\\
151	2.30258388519287\\
152	2.30258464813232\\
153	2.30258555412292\\
154	2.30258402824402\\
155	2.30258502960205\\
156	2.30258493423462\\
157	2.30258402824402\\
158	2.30258474349976\\
159	2.30258522033691\\
160	2.30258526802063\\
161	2.30258502960205\\
162	2.30258464813232\\
163	2.3025857925415\\
164	2.30258545875549\\
165	2.30258479118347\\
166	2.30258593559265\\
167	2.30258631706238\\
168	2.30258512496948\\
169	2.30258417129517\\
170	2.30258474349976\\
171	2.30258555412292\\
172	2.30258498191834\\
173	2.3025848865509\\
174	2.30258522033691\\
175	2.30258331298828\\
176	2.30258469581604\\
177	2.3025851726532\\
178	2.30258641242981\\
179	2.30258474349976\\
180	2.30258560180664\\
181	2.30258545875549\\
182	2.30258378982544\\
183	2.30258564949036\\
184	2.30258412361145\\
185	2.3025848865509\\
186	2.30258545875549\\
187	2.30258564949036\\
188	2.30258455276489\\
189	2.30258412361145\\
190	2.30258436203003\\
191	2.30258479118347\\
192	2.30258574485779\\
193	2.30258526802063\\
194	2.30258593559265\\
195	2.3025857925415\\
196	2.30258493423462\\
197	2.30258445739746\\
198	2.30258412361145\\
199	2.3025839805603\\
200	2.30258436203003\\
201	2.3025848865509\\
202	2.30258474349976\\
203	2.30258526802063\\
204	2.30258412361145\\
205	2.30258431434631\\
206	2.30258417129517\\
207	2.30258388519287\\
208	2.30258522033691\\
209	2.30258498191834\\
210	2.30258402824402\\
211	2.3025842666626\\
212	2.30258436203003\\
213	2.30258431434631\\
214	2.30258545875549\\
215	2.30258493423462\\
216	2.3025851726532\\
217	2.30258474349976\\
218	2.30258526802063\\
219	2.30258507728577\\
220	2.30258564949036\\
221	2.30258507728577\\
222	2.30258512496948\\
223	2.30258474349976\\
224	2.3025857925415\\
225	2.30258522033691\\
226	2.30258560180664\\
227	2.3025851726532\\
228	2.30258483886719\\
229	2.30258469581604\\
230	2.30258507728577\\
231	2.30258483886719\\
232	2.30258564949036\\
233	2.3025848865509\\
234	2.30258531570435\\
235	2.30258460044861\\
236	2.30258507728577\\
237	2.30258526802063\\
238	2.30258364677429\\
239	2.30258507728577\\
240	2.30258550643921\\
241	2.30258440971375\\
242	2.30258512496948\\
243	2.30258507728577\\
244	2.30258550643921\\
245	2.30258569717407\\
246	2.30258507728577\\
247	2.30258541107178\\
248	2.3025848865509\\
249	2.30258421897888\\
250	2.30258498191834\\
251	2.30258436203003\\
};
\addlegendentry{FedAvg}

\end{axis}

\begin{axis}[%
width=6in,
height=4.8in,
at={(0in,0in)},
scale only axis,
xmin=0,
xmax=1,     
ymin=0,
ymax=1,
axis line style={draw=none},
ticks=none,
axis x line*=bottom,
axis y line*=left
]
\end{axis}

%% file: Figures/F4.tex
%
%

\definecolor{mycolor1}{rgb}{0.00000,0.44700,0.74100}%
\definecolor{mycolor2}{rgb}{0.85000,0.32500,0.09800}%
\definecolor{mycolor3}{rgb}{0.92900,0.69400,0.12500}%
\definecolor{mycolor4}{rgb}{0.49400,0.18400,0.55600}%
\definecolor{mycolor5}{rgb}{0.74902,0.00000,0.74902}%
\definecolor{mycolor6}{rgb}{0.00000,0.44706,0.74118}%
\definecolor{mycolor7}{rgb}{0.63500,0.07800,0.18400}

\begin{axis}[%
width=8in,
height=7in,
at={(0.78in,0.556in)},
scale only axis,
xmin=0,
xmax=255,
xtick={0,50,...,250},
ytick={0,0.1,...,0.9},
xlabel style={font=\color{white!15!black}},
xlabel={\Huge{$\text{Communication rounds}$}},
ymin=0,
ymax=0.9,
ylabel style={font=\color{white!15!black}},
ylabel={\Huge{$\text{Training accuracy} $}},
axis background/.style={fill=white},
xmajorgrids,
ymajorgrids,
legend style={at={(0.603,0.692)}, anchor=south west, legend cell align=left, align=left, draw=white!15!black,font=\huge}
]
\addplot [color=mycolor1, line width=3.0pt, mark repeat={20}, mark size=6.0pt, mark=o, mark options={solid, mycolor1}]
  table[row sep=crcr]{%
1	0.0970406666666667\\
2	0.130446\\
3	0.21572\\
4	0.267899333333333\\
5	0.291796666666667\\
6	0.340778666666667\\
7	0.393344666666667\\
8	0.423778666666667\\
9	0.469124\\
10	0.477264666666667\\
11	0.47421\\
12	0.500446\\
13	0.534720666666667\\
14	0.540843333333333\\
15	0.574778666666667\\
16	0.580726\\
17	0.590308666666667\\
18	0.609125333333333\\
19	0.630708666666667\\
20	0.630076666666667\\
21	0.643320666666667\\
22	0.644476666666667\\
23	0.659581333333333\\
24	0.662648666666667\\
25	0.648738666666667\\
26	0.664056666666667\\
27	0.683414666666667\\
28	0.689673333333333\\
29	0.68111\\
30	0.694142\\
31	0.694829333333333\\
32	0.694269333333333\\
33	0.694096\\
34	0.686071333333333\\
35	0.707566\\
36	0.7237\\
37	0.71851\\
38	0.724440666666667\\
39	0.721865333333333\\
40	0.734508\\
41	0.735814666666667\\
42	0.726728666666667\\
43	0.731148666666667\\
44	0.739781333333333\\
45	0.744124666666667\\
46	0.746399333333333\\
47	0.746226666666667\\
48	0.742066\\
49	0.755080666666667\\
50	0.754283333333333\\
51	0.753020666666667\\
52	0.752111333333333\\
53	0.75531\\
54	0.756258\\
55	0.756151333333333\\
56	0.761726666666667\\
57	0.767462\\
58	0.767205333333333\\
59	0.768491333333333\\
60	0.770033333333334\\
61	0.770404\\
62	0.766433333333333\\
63	0.768567333333333\\
64	0.76738\\
65	0.771563333333333\\
66	0.772430666666667\\
67	0.776412\\
68	0.774396\\
69	0.77786\\
70	0.779384\\
71	0.782072666666667\\
72	0.777002\\
73	0.778192666666667\\
74	0.779424666666666\\
75	0.781565333333333\\
76	0.786132666666666\\
77	0.787267333333333\\
78	0.786861333333333\\
79	0.788892666666667\\
80	0.789986\\
81	0.792262\\
82	0.791802\\
83	0.794611333333333\\
84	0.794744\\
85	0.791873333333333\\
86	0.796815333333333\\
87	0.79729\\
88	0.792932666666667\\
89	0.794141333333333\\
90	0.794584666666667\\
91	0.795985333333333\\
92	0.799678\\
93	0.798694\\
94	0.799107333333333\\
95	0.799166\\
96	0.800128666666667\\
97	0.800262666666667\\
98	0.800002\\
99	0.802704666666667\\
100	0.801446666666667\\
101	0.803465333333334\\
102	0.804111333333333\\
103	0.803402\\
104	0.805065333333333\\
105	0.806156\\
106	0.804942666666667\\
107	0.804411333333333\\
108	0.805374666666667\\
109	0.806936666666667\\
110	0.807705333333333\\
111	0.806866\\
112	0.808407333333333\\
113	0.807991333333333\\
114	0.80883\\
115	0.809507333333333\\
116	0.808736\\
117	0.807726\\
118	0.808134\\
119	0.809476\\
120	0.810256\\
121	0.810954666666667\\
122	0.810572666666667\\
123	0.809176666666667\\
124	0.812327333333333\\
125	0.811746666666667\\
126	0.812195333333334\\
127	0.813948666666667\\
128	0.813634666666667\\
129	0.815784666666667\\
130	0.816383333333333\\
131	0.816896666666667\\
132	0.817398666666667\\
133	0.816415333333334\\
134	0.815794666666667\\
135	0.816179333333333\\
136	0.816037333333333\\
137	0.814954666666667\\
138	0.815847333333333\\
139	0.816216\\
140	0.816796\\
141	0.816436\\
142	0.817344\\
143	0.818246666666667\\
144	0.817996\\
145	0.819114\\
146	0.819860666666667\\
147	0.819377333333333\\
148	0.819649333333333\\
149	0.820915333333333\\
150	0.821519333333333\\
151	0.822338\\
152	0.822706666666667\\
153	0.822841333333333\\
154	0.822637333333333\\
155	0.822828\\
156	0.822512666666667\\
157	0.822534666666667\\
158	0.821868\\
159	0.822322\\
160	0.823119333333333\\
161	0.82231\\
162	0.822222\\
163	0.822531333333333\\
164	0.823872\\
165	0.824271333333333\\
166	0.824255333333333\\
167	0.823282\\
168	0.823064666666667\\
169	0.824103333333334\\
170	0.823684\\
171	0.823328\\
172	0.822507333333333\\
173	0.823525333333334\\
174	0.825159333333333\\
175	0.825299333333333\\
176	0.824784\\
177	0.823903333333333\\
178	0.825241333333333\\
179	0.826362666666666\\
180	0.825975333333333\\
181	0.825576\\
182	0.825548666666667\\
183	0.825373333333333\\
184	0.825981333333333\\
185	0.825986\\
186	0.826469333333334\\
187	0.826520666666667\\
188	0.827033333333333\\
189	0.827503333333333\\
190	0.827567333333333\\
191	0.827268\\
192	0.827018\\
193	0.826301333333333\\
194	0.826565333333333\\
195	0.827707333333333\\
196	0.82741\\
197	0.82792\\
198	0.829205333333334\\
199	0.829996666666667\\
200	0.829924\\
201	0.829506666666667\\
202	0.829940666666667\\
203	0.829004666666667\\
204	0.829332\\
205	0.829335333333333\\
206	0.828780666666667\\
207	0.828177333333333\\
208	0.828307333333333\\
209	0.829511333333333\\
210	0.829535333333333\\
211	0.830734666666667\\
212	0.831715333333333\\
213	0.831274\\
214	0.831332\\
215	0.831398666666667\\
216	0.831922\\
217	0.831236666666667\\
218	0.830757333333333\\
219	0.831663333333333\\
220	0.831562\\
221	0.83142\\
222	0.831307333333334\\
223	0.831754\\
224	0.831923333333333\\
225	0.831782666666666\\
226	0.831357333333333\\
227	0.831814666666666\\
228	0.831814\\
229	0.832157333333333\\
230	0.83229\\
231	0.83192\\
232	0.83203\\
233	0.832044666666667\\
234	0.831922\\
235	0.832462\\
236	0.832674\\
237	0.832934\\
238	0.833365333333333\\
239	0.832924666666667\\
240	0.833681333333333\\
241	0.8342\\
242	0.834084666666667\\
243	0.834390666666667\\
244	0.834513333333333\\
245	0.833797333333333\\
246	0.833526\\
247	0.833842666666667\\
248	0.833844666666667\\
249	0.834138\\
250	0.834314\\
251	0.834546666666667\\
};
\addlegendentry{Joint}

\addplot [color=mycolor2, line width=3.0pt, mark repeat={20}, mark size=6.0pt, mark=square, mark options={solid, mycolor2}]
  table[row sep=crcr]{%
1	0.0901366666666667\\
2	0.115715\\
3	0.127191666666667\\
4	0.14228\\
5	0.158098333333333\\
6	0.171151666666667\\
7	0.180585\\
8	0.196295\\
9	0.207171666666667\\
10	0.21757\\
11	0.229173333333333\\
12	0.241401666666667\\
13	0.256881666666667\\
14	0.269258333333333\\
15	0.280938333333333\\
16	0.288906666666667\\
17	0.295488333333333\\
18	0.307033333333333\\
19	0.310491666666667\\
20	0.316525\\
21	0.324761666666667\\
22	0.332343333333333\\
23	0.339915\\
24	0.344656666666667\\
25	0.352138333333333\\
26	0.361996666666667\\
27	0.365645\\
28	0.373696666666667\\
29	0.377248333333333\\
30	0.383846666666667\\
31	0.386338333333333\\
32	0.390846666666667\\
33	0.399805\\
34	0.404061666666667\\
35	0.409731666666667\\
36	0.412646666666667\\
37	0.416501666666667\\
38	0.421663333333333\\
39	0.423921666666667\\
40	0.426143333333333\\
41	0.428355\\
42	0.431218333333333\\
43	0.433796666666667\\
44	0.439376666666667\\
45	0.442341666666667\\
46	0.443255\\
47	0.444081666666667\\
48	0.447641666666667\\
49	0.449381666666667\\
50	0.451938333333333\\
51	0.455476666666667\\
52	0.45892\\
53	0.46241\\
54	0.465186666666667\\
55	0.468471666666667\\
56	0.470453333333333\\
57	0.473486666666667\\
58	0.474471666666667\\
59	0.475228333333333\\
60	0.475748333333333\\
61	0.475676666666667\\
62	0.479633333333333\\
63	0.480876666666667\\
64	0.482918333333333\\
65	0.485925\\
66	0.48645\\
67	0.488335\\
68	0.488638333333333\\
69	0.489251666666667\\
70	0.490653333333333\\
71	0.490548333333333\\
72	0.492301666666667\\
73	0.494581666666667\\
74	0.497013333333333\\
75	0.498428333333333\\
76	0.500021666666666\\
77	0.501715\\
78	0.502871666666667\\
79	0.504443333333333\\
80	0.505763333333333\\
81	0.507455\\
82	0.508788333333333\\
83	0.510338333333333\\
84	0.510561666666667\\
85	0.511903333333333\\
86	0.513246666666667\\
87	0.51346\\
88	0.514866666666667\\
89	0.514956666666667\\
90	0.51591\\
91	0.515736666666667\\
92	0.517633333333333\\
93	0.519048333333333\\
94	0.519716666666667\\
95	0.521248333333333\\
96	0.521593333333333\\
97	0.521698333333333\\
98	0.523561666666667\\
99	0.524666666666667\\
100	0.525385\\
101	0.526021666666667\\
102	0.527876666666667\\
103	0.528398333333333\\
104	0.528581666666667\\
105	0.529675\\
106	0.530233333333333\\
107	0.53112\\
108	0.532171666666667\\
109	0.532093333333333\\
110	0.533523333333333\\
111	0.534796666666667\\
112	0.53651\\
113	0.536373333333333\\
114	0.537073333333333\\
115	0.538698333333333\\
116	0.53911\\
117	0.539826666666667\\
118	0.541496666666667\\
119	0.541881666666667\\
120	0.54316\\
121	0.544138333333333\\
122	0.54485\\
123	0.545123333333333\\
124	0.545186666666667\\
125	0.546061666666667\\
126	0.54772\\
127	0.547805\\
128	0.548701666666667\\
129	0.549243333333333\\
130	0.550036666666667\\
131	0.550123333333333\\
132	0.551321666666667\\
133	0.552033333333333\\
134	0.552503333333333\\
135	0.55312\\
136	0.553888333333333\\
137	0.55427\\
138	0.555751666666667\\
139	0.556071666666667\\
140	0.5571\\
141	0.557641666666667\\
142	0.55783\\
143	0.55843\\
144	0.558916666666667\\
145	0.559386666666667\\
146	0.559793333333333\\
147	0.560311666666667\\
148	0.560831666666667\\
149	0.56168\\
150	0.562181666666667\\
151	0.56273\\
152	0.563108333333333\\
153	0.563671666666667\\
154	0.564038333333333\\
155	0.564295\\
156	0.564065\\
157	0.564113333333333\\
158	0.564858333333333\\
159	0.565276666666667\\
160	0.565968333333333\\
161	0.567196666666667\\
162	0.567901666666667\\
163	0.568051666666667\\
164	0.568665\\
165	0.569116666666667\\
166	0.569021666666667\\
167	0.569333333333333\\
168	0.569756666666667\\
169	0.570091666666667\\
170	0.570928333333333\\
171	0.571433333333333\\
172	0.571673333333333\\
173	0.57247\\
174	0.57331\\
175	0.573586666666667\\
176	0.574235\\
177	0.574728333333333\\
178	0.574753333333333\\
179	0.575175\\
180	0.575695\\
181	0.576215\\
182	0.576801666666667\\
183	0.57764\\
184	0.578596666666667\\
185	0.578436666666667\\
186	0.579166666666667\\
187	0.57933\\
188	0.579643333333333\\
189	0.579736666666667\\
190	0.580358333333333\\
191	0.580346666666667\\
192	0.581238333333333\\
193	0.581628333333333\\
194	0.581123333333333\\
195	0.58226\\
196	0.582478333333333\\
197	0.58277\\
198	0.582858333333333\\
199	0.583218333333333\\
200	0.58338\\
201	0.583276666666667\\
202	0.583948333333333\\
203	0.583645\\
204	0.584041666666667\\
205	0.58429\\
206	0.584091666666667\\
207	0.583951666666667\\
208	0.584408333333333\\
209	0.58472\\
210	0.584918333333333\\
211	0.584996666666667\\
212	0.58533\\
213	0.585475\\
214	0.58575\\
215	0.585973333333333\\
216	0.58671\\
217	0.586681666666667\\
218	0.586745\\
219	0.586695\\
220	0.587105\\
221	0.587138333333333\\
222	0.587435\\
223	0.587688333333333\\
224	0.588091666666667\\
225	0.588738333333333\\
226	0.58915\\
227	0.5899\\
228	0.589823333333333\\
229	0.590225\\
230	0.590728333333334\\
231	0.591093333333333\\
232	0.591588333333333\\
233	0.591763333333333\\
234	0.591755\\
235	0.591338333333334\\
236	0.591403333333333\\
237	0.591605\\
238	0.591833333333333\\
239	0.591856666666667\\
240	0.592345\\
241	0.592418333333333\\
242	0.592435\\
243	0.592805\\
244	0.593033333333333\\
245	0.593551666666667\\
246	0.593678333333333\\
247	0.593988333333333\\
248	0.594231666666667\\
249	0.594575\\
250	0.59504\\
251	0.595206666666667\\
};
\addlegendentry{UL-only}

\addplot [color=mycolor3, line width=3.0pt, mark repeat={20}, mark size=6.0pt, mark=diamond, mark options={solid, mycolor3}]
  table[row sep=crcr]{%
1	0.0901933333333333\\
2	0.09888\\
3	0.102926666666667\\
4	0.0997733333333333\\
5	0.100056666666667\\
6	0.09817\\
7	0.09996\\
8	0.09806\\
9	0.103903333333333\\
10	0.098\\
11	0.101853333333333\\
12	0.102446666666667\\
13	0.101066666666667\\
14	0.101536666666667\\
15	0.0992233333333334\\
16	0.0976866666666667\\
17	0.10062\\
18	0.0964766666666667\\
19	0.10221\\
20	0.10415\\
21	0.09693\\
22	0.099\\
23	0.09957\\
24	0.0985033333333333\\
25	0.09772\\
26	0.10018\\
27	0.101896666666667\\
28	0.107776666666667\\
29	0.10644\\
30	0.104403333333333\\
31	0.10179\\
32	0.102223333333333\\
33	0.103486666666667\\
34	0.103196666666667\\
35	0.0989466666666667\\
36	0.0982033333333333\\
37	0.0997733333333333\\
38	0.0983866666666667\\
39	0.09924\\
40	0.10376\\
41	0.10304\\
42	0.100073333333333\\
43	0.101863333333333\\
44	0.10503\\
45	0.102433333333333\\
46	0.09729\\
47	0.0996033333333333\\
48	0.10247\\
49	0.10257\\
50	0.102746666666667\\
51	0.100516666666667\\
52	0.0993\\
53	0.0994766666666667\\
54	0.100656666666667\\
55	0.102313333333333\\
56	0.0984833333333333\\
57	0.0970633333333333\\
58	0.102686666666667\\
59	0.101633333333333\\
60	0.101173333333333\\
61	0.0948433333333333\\
62	0.103263333333333\\
63	0.10406\\
64	0.09974\\
65	0.102136666666667\\
66	0.0986633333333333\\
67	0.0983066666666667\\
68	0.104986666666667\\
69	0.103216666666667\\
70	0.103726666666667\\
71	0.103726666666667\\
72	0.101013333333333\\
73	0.100843333333333\\
74	0.100103333333333\\
75	0.104363333333333\\
76	0.101783333333333\\
77	0.102433333333333\\
78	0.0981733333333333\\
79	0.0974966666666667\\
80	0.0963566666666667\\
81	0.10035\\
82	0.102993333333333\\
83	0.09908\\
84	0.10116\\
85	0.09865\\
86	0.102416666666667\\
87	0.0980433333333333\\
88	0.102933333333333\\
89	0.101053333333333\\
90	0.102703333333333\\
91	0.0986633333333333\\
92	0.0983633333333333\\
93	0.10109\\
94	0.0983433333333334\\
95	0.09745\\
96	0.10365\\
97	0.0979533333333333\\
98	0.0974966666666667\\
99	0.0972566666666667\\
100	0.09897\\
101	0.102046666666667\\
102	0.102313333333333\\
103	0.100026666666667\\
104	0.101096666666667\\
105	0.0994166666666667\\
106	0.0993266666666667\\
107	0.100396666666667\\
108	0.103796666666667\\
109	0.100543333333333\\
110	0.0995833333333333\\
111	0.102103333333333\\
112	0.09955\\
113	0.0983633333333333\\
114	0.100313333333333\\
115	0.09958\\
116	0.100226666666667\\
117	0.0992866666666667\\
118	0.100073333333333\\
119	0.0982366666666667\\
120	0.100626666666667\\
121	0.10038\\
122	0.0991333333333333\\
123	0.102346666666667\\
124	0.1039\\
125	0.103726666666667\\
126	0.09819\\
127	0.100456666666667\\
128	0.1023\\
129	0.09699\\
130	0.10301\\
131	0.101336666666667\\
132	0.09368\\
133	0.100656666666667\\
134	0.10004\\
135	0.107006666666667\\
136	0.0981766666666667\\
137	0.104506666666667\\
138	0.10045\\
139	0.104506666666667\\
140	0.0999333333333334\\
141	0.0968433333333333\\
142	0.09689\\
143	0.103976666666667\\
144	0.0993566666666667\\
145	0.101146666666667\\
146	0.0968733333333333\\
147	0.09886\\
148	0.10264\\
149	0.101956666666667\\
150	0.101336666666667\\
151	0.1053\\
152	0.10135\\
153	0.09958\\
154	0.103626666666667\\
155	0.101143333333333\\
156	0.101766666666667\\
157	0.100666666666667\\
158	0.0989166666666666\\
159	0.0997033333333333\\
160	0.1023\\
161	0.0977533333333334\\
162	0.101853333333333\\
163	0.0969933333333333\\
164	0.0992866666666667\\
165	0.0996833333333333\\
166	0.0994166666666667\\
167	0.0967533333333333\\
168	0.102386666666667\\
169	0.102123333333333\\
170	0.0974333333333333\\
171	0.09689\\
172	0.09716\\
173	0.0999633333333334\\
174	0.0990366666666667\\
175	0.10092\\
176	0.106186666666667\\
177	0.10038\\
178	0.0953533333333333\\
179	0.09794\\
180	0.0986833333333333\\
181	0.10235\\
182	0.100156666666667\\
183	0.101986666666667\\
184	0.103543333333333\\
185	0.0964866666666667\\
186	0.100513333333333\\
187	0.09841\\
188	0.105283333333333\\
189	0.101716666666667\\
190	0.10389\\
191	0.102746666666667\\
192	0.10344\\
193	0.1005\\
194	0.09759\\
195	0.10006\\
196	0.101293333333333\\
197	0.104496666666667\\
198	0.102416666666667\\
199	0.10194\\
200	0.103486666666667\\
201	0.104783333333333\\
202	0.10301\\
203	0.100636666666667\\
204	0.11033\\
205	0.100403333333333\\
206	0.102926666666667\\
207	0.10409\\
208	0.102283333333333\\
209	0.09853\\
210	0.101246666666667\\
211	0.103036666666667\\
212	0.10153\\
213	0.102343333333333\\
214	0.10003\\
215	0.09788\\
216	0.101116666666667\\
217	0.106873333333333\\
218	0.09841\\
219	0.102773333333333\\
220	0.0985433333333333\\
221	0.09794\\
222	0.101516666666667\\
223	0.102703333333333\\
224	0.0969933333333333\\
225	0.09745\\
226	0.1019\\
227	0.102553333333333\\
228	0.100826666666667\\
229	0.10197\\
230	0.0996033333333333\\
231	0.0991633333333334\\
232	0.0982933333333333\\
233	0.100543333333333\\
234	0.10243\\
235	0.101306666666667\\
236	0.10163\\
237	0.100843333333333\\
238	0.105046666666667\\
239	0.10188\\
240	0.0993433333333333\\
241	0.10318\\
242	0.10142\\
243	0.1\\
244	0.10214\\
245	0.09756\\
246	0.0982833333333333\\
247	0.09892\\
248	0.10098\\
249	0.10176\\
250	0.10045\\
251	0.101746666666667\\
};
\addlegendentry{FedAvg}

\end{axis}

\begin{axis}[%
width=6in,
height=4.8in,
at={(0in,0in)},
scale only axis,
xmin=0,
xmax=1,
ymin=0,
ymax=1,
axis line style={draw=none},
ticks=none,
axis x line*=bottom,
axis y line*=left
]
\end{axis}

%% file: Figures/DiffE.tex
%
%

\definecolor{mycolor1}{rgb}{0.00000,0.44700,0.74100}%
\definecolor{mycolor2}{rgb}{0.85000,0.32500,0.09800}%
\definecolor{mycolor3}{rgb}{0.92900,0.69400,0.12500}%
\definecolor{mycolor4}{rgb}{0.49400,0.18400,0.55600}%
\definecolor{mycolor5}{rgb}{0.74902,0.00000,0.74902}%
\definecolor{mycolor6}{rgb}{0.00000,0.44706,0.74118}%
\definecolor{mycolor7}{rgb}{0.63500,0.07800,0.18400}


\begin{axis}[%
width=7.9in,
height=6.8in,
at={(0.78in,0.57in)},
scale only axis,
bar shift auto,
xmin=0.514285714285714,
xmax=7.48571428571429,
xtick={1,2,3,4,5,6,7},
xticklabels={{1},{2},{3},{4},{5},{7},{10}},
xlabel style={font=\color{white!15!black}},
xlabel={\Huge{$E$}},
ymin=0,
ymax=0.9,
ytick = {0,0.1,...,0.9},
ylabel style={font=\color{white!15!black}},
ylabel={\Huge{$\text{Testing accuracy} $}},
axis background/.style={fill=white},
xmajorgrids,
ymajorgrids,
legend style={at={(0.612,0.762)}, anchor=south west, legend cell align=left, align=left, draw=white!15!black,font=\huge}
]
\addplot[ybar, bar width=0.229, fill=mycolor1, draw=black, area legend] table[row sep=crcr] {%
1	0.7856\\
2	0.8419\\
3	0.7377\\
4	0.7123\\
5	0.4934\\
6	0.3849\\
7	0.107\\
};
\addplot[forget plot, color=white!15!black, line width=0.8pt] table[row sep=crcr] {%
0.514285714285714	0\\
7.48571428571429	0\\
};
\addlegendentry{Joint}

\addplot[ybar, bar width=0.229, fill=mycolor2, draw=black, area legend] table[row sep=crcr] {%
1	0.4507\\
2	0.6001\\
3	0.3446\\
4	0.4036\\
5	0.4372\\
6	0.3226\\
7	0.098\\
};
\addplot[forget plot, color=white!15!black, line width=0.8pt] table[row sep=crcr] {%
0.514285714285714	0\\
7.48571428571429	0\\
};
\addlegendentry{UL-only}

\end{axis}

\begin{axis}[%
width=6in,
height=4.8in,
at={(0in,0in)},
scale only axis,
xmin=0,
xmax=1,
ymin=0,
ymax=1,
axis line style={draw=none},
ticks=none,
axis x line*=bottom,
axis y line*=left
]
\end{axis}

%% file: Figures/F3_new.tex
%
%

\definecolor{mycolor1}{rgb}{0.00000,0.44700,0.74100}%
\definecolor{mycolor2}{rgb}{0.85000,0.32500,0.09800}%
\definecolor{mycolor3}{rgb}{0.92900,0.69400,0.12500}%
\definecolor{mycolor4}{rgb}{0.49400,0.18400,0.55600}%
\definecolor{mycolor5}{rgb}{0.74902,0.00000,0.74902}%
\definecolor{mycolor6}{rgb}{0.00000,0.44706,0.74118}%
\definecolor{mycolor7}{rgb}{0.63500,0.07800,0.18400}
\definecolor{mycolor8}{rgb}{0.46667,0.67451,0.18824}%

\begin{axis}[%
width=8in,
height=7in,
at={(0.78in,0.556in)},
scale only axis,
xmin=0,
xmax=255,
xtick={0,50,...,250},
ytick={0,0.1,...,0.9},
xlabel style={font=\color{white!15!black}},
xlabel={\Huge{$\text{Communication rounds}$}},
ymin=0,
ymax=0.9,
ylabel style={font=\color{white!15!black}},
ylabel={\Huge{$\text{Training accuracy} $}},
axis background/.style={fill=white},
xmajorgrids,
ymajorgrids,
legend style={at={(0.619,0.632)}, anchor=south west, legend cell align=left, align=left, draw=white!15!black,font=\huge}
]
\addplot [color=mycolor1, line width=3.0pt, mark repeat={20}, mark size=6.0pt, mark=o, mark options={solid, mycolor1}]
  table[row sep=crcr]{%
1	0.103756666666667\\
2	0.166229444444444\\
3	0.25398\\
4	0.327144\\
5	0.392924\\
6	0.438730333333333\\
7	0.472944\\
8	0.498407333333333\\
9	0.520379\\
10	0.549556\\
11	0.57679\\
12	0.599744\\
13	0.624868\\
14	0.644236333333333\\
15	0.659599\\
16	0.674985\\
17	0.688547\\
18	0.697370333333333\\
19	0.708933333333333\\
20	0.717392\\
21	0.724759333333333\\
22	0.730099\\
23	0.734057666666667\\
24	0.734864666666667\\
25	0.737815\\
26	0.738301\\
27	0.741396\\
28	0.744452\\
29	0.745840666666667\\
30	0.748896\\
31	0.753064333333333\\
32	0.754511\\
33	0.758283333333334\\
34	0.763375333333334\\
35	0.764602666666667\\
36	0.766547333333333\\
37	0.769959666666667\\
38	0.772334666666667\\
39	0.773351666666667\\
40	0.776351\\
41	0.778089\\
42	0.780006333333333\\
43	0.780806333333333\\
44	0.784789666666667\\
45	0.785951333333333\\
46	0.786549666666667\\
47	0.785702333333333\\
48	0.786597333333333\\
49	0.784862333333333\\
50	0.785101666666667\\
51	0.787147333333333\\
52	0.791127\\
53	0.79416\\
54	0.797487\\
55	0.800586666666667\\
56	0.801169333333334\\
57	0.800849666666667\\
58	0.799273333333333\\
59	0.799057\\
60	0.799609333333333\\
61	0.801178333333333\\
62	0.803146666666667\\
63	0.807065\\
64	0.8093\\
65	0.81025\\
66	0.811168666666667\\
67	0.81177\\
68	0.810846333333334\\
69	0.810196333333334\\
70	0.810531\\
71	0.810843666666667\\
72	0.810883\\
73	0.811426666666667\\
74	0.810971333333333\\
75	0.811516\\
76	0.812091333333333\\
77	0.811806333333334\\
78	0.812427\\
79	0.814686333333333\\
80	0.813149\\
81	0.812173666666667\\
82	0.813319\\
83	0.813549333333333\\
84	0.813922\\
85	0.817464666666667\\
86	0.819821333333333\\
87	0.820277333333333\\
88	0.821191666666667\\
89	0.821450333333334\\
90	0.821098\\
91	0.822060666666667\\
92	0.823591666666667\\
93	0.824687666666667\\
94	0.825893\\
95	0.826553333333333\\
96	0.827039333333333\\
97	0.826753333333333\\
98	0.826922333333333\\
99	0.826825333333333\\
100	0.826785333333333\\
101	0.825973666666667\\
102	0.825898666666667\\
103	0.825651\\
104	0.825443333333334\\
105	0.825810666666667\\
106	0.826628333333333\\
107	0.827016\\
108	0.826918333333333\\
109	0.826275\\
110	0.825638\\
111	0.824666\\
112	0.825094333333333\\
113	0.824734666666667\\
114	0.826260666666667\\
115	0.828780666666667\\
116	0.829744666666667\\
117	0.830667666666667\\
118	0.831350333333333\\
119	0.830035333333333\\
120	0.828117666666667\\
121	0.828399\\
122	0.827724\\
123	0.828517\\
124	0.829758\\
125	0.830847333333333\\
126	0.831417333333333\\
127	0.831689333333333\\
128	0.831964\\
129	0.832433\\
130	0.832410666666667\\
131	0.832368666666667\\
132	0.832872\\
133	0.833632\\
134	0.834123\\
135	0.834846\\
136	0.835475666666667\\
137	0.835695666666667\\
138	0.835781666666667\\
139	0.836155666666667\\
140	0.836698\\
141	0.837387\\
142	0.838110666666667\\
143	0.838603333333333\\
144	0.838782333333333\\
145	0.838702666666667\\
146	0.838695\\
147	0.838712666666667\\
148	0.838608666666667\\
149	0.838534\\
150	0.838748\\
151	0.838990666666667\\
152	0.839034\\
153	0.839084333333333\\
154	0.838947666666667\\
155	0.838832\\
156	0.838284333333333\\
157	0.838204333333333\\
158	0.838443333333333\\
159	0.839010333333333\\
160	0.839463\\
161	0.840140666666667\\
162	0.840616\\
163	0.840941333333333\\
164	0.840659333333333\\
165	0.840429666666667\\
166	0.840291\\
167	0.840264666666667\\
168	0.840271\\
169	0.840642\\
170	0.841151\\
171	0.841708333333333\\
172	0.841803333333333\\
173	0.841859333333333\\
174	0.842043666666667\\
175	0.84218\\
176	0.842348666666667\\
177	0.842834666666667\\
178	0.843186666666667\\
179	0.843427666666667\\
180	0.843385666666667\\
181	0.843427333333333\\
182	0.843511333333333\\
183	0.843446\\
184	0.843348\\
185	0.843346\\
186	0.843338666666667\\
187	0.843428333333333\\
188	0.843649\\
189	0.84372\\
190	0.843799666666667\\
191	0.843371666666667\\
192	0.842905\\
193	0.842828666666667\\
194	0.842837666666667\\
195	0.842927333333333\\
196	0.843529333333333\\
197	0.843931\\
198	0.844103333333333\\
199	0.844447333333333\\
200	0.843663333333333\\
201	0.84376\\
202	0.843401666666667\\
203	0.843299\\
204	0.843455666666667\\
205	0.844756666666667\\
206	0.845289\\
207	0.846504333333333\\
208	0.845625666666667\\
209	0.844562333333333\\
210	0.84352\\
211	0.842086\\
212	0.840510666666667\\
213	0.840911\\
214	0.84098\\
215	0.840967666666667\\
216	0.841037333333333\\
217	0.841120333333333\\
218	0.841195\\
219	0.841808666666667\\
220	0.842539666666667\\
221	0.843373\\
222	0.843990333333333\\
223	0.844443\\
224	0.844611333333333\\
225	0.844691333333333\\
226	0.843564\\
227	0.843978666666667\\
228	0.844058666666667\\
229	0.842467\\
230	0.843105333333333\\
231	0.843971666666667\\
232	0.844749666666666\\
233	0.844643666666667\\
234	0.847015\\
235	0.845585666666667\\
236	0.845554333333333\\
237	0.844572666666667\\
238	0.845214333333334\\
239	0.843563666666667\\
240	0.843596\\
241	0.843273666666667\\
242	0.842745333333333\\
243	0.841926666666667\\
244	0.843242333333333\\
245	0.844345666666667\\
246	0.845222333333333\\
247	0.846116\\
248	0.846941333333333\\
249	0.847154666666667\\
250	0.847495\\
251	0.847615\\
};
\addlegendentry{h = 25 m}

\addplot [color=mycolor4, line width=3.0pt, mark repeat={20}, mark size=6.0pt, mark=o, mark options={solid, mycolor4}]
  table[row sep=crcr]{%
1	0.105078333333333\\
2	0.205268333333333\\
3	0.389728333333333\\
4	0.454448333333333\\
5	0.511611666666667\\
6	0.557823333333333\\
7	0.63219\\
8	0.65624\\
9	0.680958333333333\\
10	0.689661666666667\\
11	0.701268333333333\\
12	0.723983333333333\\
13	0.729043333333333\\
14	0.735985\\
15	0.75471\\
16	0.760801666666667\\
17	0.762473333333333\\
18	0.760145\\
19	0.765575\\
20	0.76669\\
21	0.772018333333333\\
22	0.7772\\
23	0.776726666666667\\
24	0.782045\\
25	0.78313\\
26	0.788883333333333\\
27	0.790263333333333\\
28	0.79383\\
29	0.798893333333333\\
30	0.798478333333333\\
31	0.801236666666666\\
32	0.803405\\
33	0.803615\\
34	0.805448333333333\\
35	0.808603333333333\\
36	0.809005\\
37	0.808771666666667\\
38	0.80942\\
39	0.80958\\
40	0.810266666666667\\
41	0.811581666666667\\
42	0.81207\\
43	0.812113333333333\\
44	0.811655\\
45	0.811888333333333\\
46	0.81423\\
47	0.81627\\
48	0.817778333333333\\
49	0.816981666666667\\
50	0.817343333333333\\
51	0.816918333333333\\
52	0.81825\\
53	0.818648333333333\\
54	0.821196666666667\\
55	0.820865\\
56	0.822528333333333\\
57	0.822236666666667\\
58	0.823133333333333\\
59	0.822783333333333\\
60	0.823093333333333\\
61	0.825256666666667\\
62	0.824546666666667\\
63	0.825528333333333\\
64	0.826231666666666\\
65	0.826395\\
66	0.827006666666667\\
67	0.828278333333333\\
68	0.828263333333333\\
69	0.828608333333333\\
70	0.828761666666667\\
71	0.829376666666667\\
72	0.830486666666667\\
73	0.830563333333333\\
74	0.830933333333333\\
75	0.831368333333333\\
76	0.832573333333333\\
77	0.83291\\
78	0.832916666666667\\
79	0.832621666666667\\
80	0.833303333333333\\
81	0.833235\\
82	0.834043333333333\\
83	0.83484\\
84	0.83486\\
85	0.835125\\
86	0.835521666666666\\
87	0.835371666666667\\
88	0.835775\\
89	0.835473333333333\\
90	0.835343333333333\\
91	0.835758333333333\\
92	0.836703333333333\\
93	0.836693333333333\\
94	0.836735\\
95	0.837241666666667\\
96	0.836345\\
97	0.83658\\
98	0.836798333333334\\
99	0.836608333333334\\
100	0.836795\\
101	0.837643333333333\\
102	0.837613333333333\\
103	0.837363333333333\\
104	0.837833333333333\\
105	0.838278333333334\\
106	0.838848333333333\\
107	0.83903\\
108	0.839841666666667\\
109	0.839946666666667\\
110	0.840561666666667\\
111	0.84017\\
112	0.84046\\
113	0.840848333333333\\
114	0.841143333333333\\
115	0.840965\\
116	0.840475\\
117	0.84116\\
118	0.841438333333333\\
119	0.841265\\
120	0.841593333333333\\
121	0.841695\\
122	0.841658333333333\\
123	0.841358333333333\\
124	0.841618333333333\\
125	0.841851666666667\\
126	0.842168333333333\\
127	0.842781666666667\\
128	0.843415\\
129	0.843426666666667\\
130	0.84358\\
131	0.843715\\
132	0.843951666666667\\
133	0.844241666666667\\
134	0.844243333333333\\
135	0.843963333333333\\
136	0.844838333333333\\
137	0.844745\\
138	0.845046666666667\\
139	0.845148333333333\\
140	0.845521666666667\\
141	0.845691666666667\\
142	0.846126666666667\\
143	0.846141666666667\\
144	0.846335\\
145	0.846413333333333\\
146	0.846316666666667\\
147	0.846621666666667\\
148	0.846783333333333\\
149	0.846826666666667\\
150	0.846943333333333\\
151	0.847208333333333\\
152	0.847073333333333\\
153	0.846658333333333\\
154	0.847026666666667\\
155	0.847185\\
156	0.847428333333333\\
157	0.847618333333333\\
158	0.847335\\
159	0.847618333333333\\
160	0.847581666666666\\
161	0.848078333333333\\
162	0.848336666666667\\
163	0.848473333333333\\
164	0.84883\\
165	0.848745\\
166	0.848795\\
167	0.848915\\
168	0.849123333333333\\
169	0.849206666666667\\
170	0.849488333333333\\
171	0.849501666666667\\
172	0.849456666666667\\
173	0.849525\\
174	0.849703333333333\\
175	0.849471666666667\\
176	0.8493\\
177	0.849571666666667\\
178	0.849473333333333\\
179	0.84991\\
180	0.8501\\
181	0.850273333333333\\
182	0.850508333333333\\
183	0.850448333333333\\
184	0.850135\\
185	0.850351666666667\\
186	0.850486666666667\\
187	0.850588333333333\\
188	0.850841666666666\\
189	0.850761666666667\\
190	0.85081\\
191	0.85089\\
192	0.851305\\
193	0.851188333333333\\
194	0.851456666666667\\
195	0.851535\\
196	0.851601666666667\\
197	0.851791666666667\\
198	0.851776666666667\\
199	0.851861666666667\\
200	0.851976666666667\\
201	0.851793333333333\\
202	0.851751666666667\\
203	0.852086666666667\\
204	0.852203333333333\\
205	0.852311666666667\\
206	0.852425\\
207	0.852336666666667\\
208	0.852471666666667\\
209	0.852603333333333\\
210	0.852683333333333\\
211	0.852905\\
212	0.853036666666667\\
213	0.852981666666667\\
214	0.852911666666667\\
215	0.853048333333333\\
216	0.852708333333334\\
217	0.852926666666667\\
218	0.852983333333333\\
219	0.85346\\
220	0.853471666666667\\
221	0.853461666666667\\
222	0.853408333333333\\
223	0.853591666666667\\
224	0.853713333333333\\
225	0.85366\\
226	0.854025\\
227	0.854135\\
228	0.854301666666667\\
229	0.854315\\
230	0.854478333333333\\
231	0.854641666666667\\
232	0.854601666666667\\
233	0.854655\\
234	0.854808333333333\\
235	0.854985\\
236	0.855118333333333\\
237	0.855178333333333\\
238	0.855248333333333\\
239	0.855313333333333\\
240	0.855223333333333\\
241	0.855335\\
242	0.85536\\
243	0.855381666666667\\
244	0.855261666666667\\
245	0.85535\\
246	0.855491666666667\\
247	0.8556\\
248	0.855645\\
249	0.855835\\
250	0.855795\\
251	0.855768333333333\\
};
\addlegendentry{h = 50 m}

\addplot [color=mycolor8, line width=3.0pt, mark repeat={20}, mark size=6.0pt, mark=o, mark options={solid, mycolor8}]
  table[row sep=crcr]{%
1	0.104283333333333\\
2	0.158574666666667\\
3	0.2071994\\
4	0.254934333333333\\
5	0.2980334\\
6	0.3412528\\
7	0.3802932\\
8	0.412265333333333\\
9	0.441401066666667\\
10	0.4670962\\
11	0.486918733333333\\
12	0.5072248\\
13	0.529521533333333\\
14	0.550622\\
15	0.570882666666667\\
16	0.5890856\\
17	0.6034416\\
18	0.614141733333333\\
19	0.622730133333333\\
20	0.629631066666667\\
21	0.636044133333333\\
22	0.640891933333333\\
23	0.645650333333333\\
24	0.650255\\
25	0.65458\\
26	0.6610612\\
27	0.669465066666667\\
28	0.676670066666667\\
29	0.683304666666667\\
30	0.690103666666667\\
31	0.694523133333333\\
32	0.697983\\
33	0.7018192\\
34	0.706045866666667\\
35	0.710034466666667\\
36	0.714517533333333\\
37	0.718215933333333\\
38	0.721312\\
39	0.7229102\\
40	0.723551666666667\\
41	0.722748266666667\\
42	0.722114933333333\\
43	0.723108133333333\\
44	0.7260096\\
45	0.730550533333333\\
46	0.736760733333333\\
47	0.742969133333333\\
48	0.747310533333333\\
49	0.750338466666667\\
50	0.751574666666667\\
51	0.751292866666667\\
52	0.750526933333333\\
53	0.750720333333333\\
54	0.7514708\\
55	0.752900266666667\\
56	0.755296466666667\\
57	0.758259266666667\\
58	0.760760866666667\\
59	0.7630358\\
60	0.7654166\\
61	0.767469933333333\\
62	0.769189933333334\\
63	0.771017333333333\\
64	0.772491066666667\\
65	0.773633266666667\\
66	0.7745472\\
67	0.775675466666667\\
68	0.776349933333333\\
69	0.777298666666667\\
70	0.7781108\\
71	0.7787322\\
72	0.7788792\\
73	0.779356\\
74	0.7795906\\
75	0.7802796\\
76	0.781448266666667\\
77	0.783051533333333\\
78	0.784626933333333\\
79	0.786632333333333\\
80	0.788363066666667\\
81	0.7898102\\
82	0.790970066666667\\
83	0.791931466666667\\
84	0.7922772\\
85	0.792247\\
86	0.791936066666667\\
87	0.791428733333334\\
88	0.7910326\\
89	0.7911252\\
90	0.791808666666667\\
91	0.793089266666667\\
92	0.794811133333333\\
93	0.7963314\\
94	0.797525066666667\\
95	0.798412933333333\\
96	0.799220866666667\\
97	0.799985066666667\\
98	0.801248066666667\\
99	0.802850266666667\\
100	0.804458466666667\\
101	0.8055876\\
102	0.806309733333334\\
103	0.806395066666667\\
104	0.8059618\\
105	0.805347133333334\\
106	0.804903\\
107	0.804673133333333\\
108	0.804786133333333\\
109	0.805199666666667\\
110	0.805829333333333\\
111	0.8064514\\
112	0.806984666666667\\
113	0.807401466666667\\
114	0.807687466666667\\
115	0.807866\\
116	0.807948466666667\\
117	0.807976\\
118	0.807890533333333\\
119	0.807689666666667\\
120	0.807445066666667\\
121	0.807327266666667\\
122	0.807339666666667\\
123	0.807603466666667\\
124	0.808160133333333\\
125	0.808810666666667\\
126	0.809539266666667\\
127	0.810393666666667\\
128	0.811191333333334\\
129	0.811967866666667\\
130	0.812834333333334\\
131	0.813692266666667\\
132	0.8143562\\
133	0.814917\\
134	0.815266733333334\\
135	0.815369733333334\\
136	0.8152958\\
137	0.815128266666667\\
138	0.814978\\
139	0.8149886\\
140	0.815144533333333\\
141	0.8154964\\
142	0.816172933333333\\
143	0.8169008\\
144	0.8175602\\
145	0.8182456\\
146	0.818799666666667\\
147	0.8190792\\
148	0.819291466666667\\
149	0.819484466666667\\
150	0.819581066666667\\
151	0.819628666666667\\
152	0.819715466666667\\
153	0.8197868\\
154	0.819857533333334\\
155	0.819981866666667\\
156	0.820138066666667\\
157	0.820203933333333\\
158	0.820264133333333\\
159	0.820231866666667\\
160	0.820134066666667\\
161	0.820029866666667\\
162	0.8201352\\
163	0.820304466666667\\
164	0.820626866666667\\
165	0.820934066666667\\
166	0.8212514\\
167	0.821429733333333\\
168	0.821590333333334\\
169	0.8217176\\
170	0.821938933333334\\
171	0.822168466666667\\
172	0.822430466666667\\
173	0.822648733333333\\
174	0.822819\\
175	0.822950733333334\\
176	0.8230862\\
177	0.823204333333333\\
178	0.823317\\
179	0.823458533333334\\
180	0.823594533333334\\
181	0.823711266666667\\
182	0.823891133333334\\
183	0.824212666666667\\
184	0.824621666666667\\
185	0.825079733333334\\
186	0.825600666666667\\
187	0.8260874\\
188	0.8264846\\
189	0.826684333333334\\
190	0.826799533333334\\
191	0.826863066666667\\
192	0.826926733333334\\
193	0.826936666666667\\
194	0.827135266666667\\
195	0.827457933333334\\
196	0.827823666666667\\
197	0.8282082\\
198	0.828712333333333\\
199	0.8291274\\
200	0.8294076\\
201	0.829596333333333\\
202	0.8296944\\
203	0.8295754\\
204	0.829324933333334\\
205	0.829021466666667\\
206	0.828639066666667\\
207	0.828245333333333\\
208	0.828024333333333\\
209	0.827985866666667\\
210	0.828072466666667\\
211	0.8283584\\
212	0.828751\\
213	0.8290214\\
214	0.829125066666667\\
215	0.829164533333334\\
216	0.829100733333333\\
217	0.828917933333333\\
218	0.828773933333333\\
219	0.828810733333333\\
220	0.828885266666667\\
221	0.829016933333333\\
222	0.8291966\\
223	0.829414466666667\\
224	0.829551333333334\\
225	0.829687533333333\\
226	0.829781733333333\\
227	0.829947266666667\\
228	0.830122333333333\\
229	0.8303084\\
230	0.830497533333333\\
231	0.830735933333333\\
232	0.830959466666667\\
233	0.8312016\\
234	0.831445266666667\\
235	0.831648333333333\\
236	0.831808866666667\\
237	0.8319748\\
238	0.832112466666667\\
239	0.832279133333333\\
240	0.832518133333333\\
241	0.832710266666667\\
242	0.8327714\\
243	0.832765666666667\\
244	0.832670733333334\\
245	0.832507266666667\\
246	0.832424533333333\\
247	0.832452066666667\\
248	0.832588222222222\\
249	0.832701222222222\\
250	0.832898037037037\\
251	0.832841666666667\\
};
\addlegendentry{h = 120 m}

\end{axis}

\begin{axis}[%
width=6in,
height=4.8in,
at={(0in,0in)},
scale only axis,
xmin=0,
xmax=1,
ymin=0,
ymax=1,
axis line style={draw=none},
ticks=none,
axis x line*=bottom,
axis y line*=left
]
\end{axis}

%% file: Figures/F11_new.tex
%
%

\definecolor{mycolor1}{rgb}{0.00000,0.44700,0.74100}%
\definecolor{mycolor2}{rgb}{0.85000,0.32500,0.09800}%
\definecolor{mycolor3}{rgb}{0.92900,0.69400,0.12500}%
\definecolor{mycolor4}{rgb}{0.49400,0.18400,0.55600}%
\definecolor{mycolor5}{rgb}{0.74902,0.00000,0.74902}%
\definecolor{mycolor6}{rgb}{0.00000,0.44706,0.74118}%
\definecolor{mycolor7}{rgb}{0.63500,0.07800,0.18400}
\definecolor{mycolor8}{rgb}{0.46667,0.67451,0.18824}%


\begin{axis}[%
width=8in,
height=7in,
at={(0.78in,0.556in)},
scale only axis,
xmin=0,
xmax=255,
xtick={0,50,...,250},
ytick={0,0.1,...,0.9},
xlabel style={font=\color{white!15!black}},
xlabel={\Huge{$\text{Communication rounds}$}},
ymin=0,
ymax=0.9,
ylabel style={font=\color{white!15!black}},
ylabel={\Huge{$\text{Training accuracy} $}},
axis background/.style={fill=white},
xmajorgrids,
ymajorgrids,
legend style={at={(0.57,0.233)}, anchor=south west, legend cell align=left, align=left, draw=white!15!black,font=\huge}
]
\addplot [color=mycolor1, line width=3.0pt, mark repeat={20}, mark size=6.0pt, mark=o, mark options={solid, mycolor1}]
table[row sep=crcr]{%
1	0.102423333333333\\
2	0.181478333333333\\
3	0.375603333333333\\
4	0.478193888888889\\
5	0.524672777777778\\
6	0.580208333333333\\
7	0.619832777777778\\
8	0.651486666666667\\
9	0.679353333333333\\
10	0.698716666666667\\
11	0.710936111111111\\
12	0.717535555555555\\
13	0.734807777777778\\
14	0.739878333333333\\
15	0.748658333333333\\
16	0.755875\\
17	0.761095555555555\\
18	0.763778888888889\\
19	0.769267777777778\\
20	0.772595555555556\\
21	0.775694444444444\\
22	0.781001111111111\\
23	0.781362777777778\\
24	0.783383333333333\\
25	0.785668333333333\\
26	0.788043888888889\\
27	0.789696666666666\\
28	0.792343888888889\\
29	0.793857222222222\\
30	0.79579\\
31	0.796011111111111\\
32	0.797043333333333\\
33	0.799736111111111\\
34	0.802468333333333\\
35	0.803712222222222\\
36	0.804811666666667\\
37	0.805102222222222\\
38	0.806792777777778\\
39	0.807625555555556\\
40	0.809203333333333\\
41	0.809828333333333\\
42	0.810130555555555\\
43	0.811793888888889\\
44	0.812358888888889\\
45	0.813193888888889\\
46	0.814151111111111\\
47	0.815133333333333\\
48	0.815612222222222\\
49	0.815998333333333\\
50	0.816646666666667\\
51	0.817808333333333\\
52	0.817535555555555\\
53	0.818722777777778\\
54	0.819257222222222\\
55	0.819408333333333\\
56	0.819605555555556\\
57	0.82051\\
58	0.821597777777778\\
59	0.822239444444444\\
60	0.822454444444444\\
61	0.823358888888889\\
62	0.824545555555556\\
63	0.82488\\
64	0.825476666666666\\
65	0.825921111111111\\
66	0.826392222222222\\
67	0.825954444444444\\
68	0.827273333333333\\
69	0.828492222222222\\
70	0.828361111111111\\
71	0.827968333333333\\
72	0.828843888888889\\
73	0.828965\\
74	0.829077222222222\\
75	0.828822777777778\\
76	0.829292777777778\\
77	0.830305\\
78	0.831012777777777\\
79	0.831915555555555\\
80	0.831889444444444\\
81	0.83169\\
82	0.831998333333333\\
83	0.832765555555556\\
84	0.833303888888889\\
85	0.833295555555555\\
86	0.833327777777778\\
87	0.833613333333333\\
88	0.833943888888889\\
89	0.834511666666666\\
90	0.834493888888889\\
91	0.834721666666667\\
92	0.835064444444444\\
93	0.835553333333333\\
94	0.835770555555555\\
95	0.835936666666667\\
96	0.835984444444444\\
97	0.836086111111111\\
98	0.836320555555556\\
99	0.836564444444444\\
100	0.836933888888889\\
101	0.837107222222222\\
102	0.837390555555556\\
103	0.837805\\
104	0.837971111111111\\
105	0.838395555555555\\
106	0.838747777777777\\
107	0.838947777777778\\
108	0.838977222222222\\
109	0.839136111111111\\
110	0.839456111111111\\
111	0.839677222222222\\
112	0.840013888888889\\
113	0.840231666666666\\
114	0.840432777777778\\
115	0.840671111111111\\
116	0.840630555555556\\
117	0.840861111111111\\
118	0.840778333333333\\
119	0.841032222222222\\
120	0.841532222222222\\
121	0.841625555555556\\
122	0.841760555555556\\
123	0.841851111111111\\
124	0.842025\\
125	0.84241\\
126	0.842402777777778\\
127	0.842625\\
128	0.842856111111111\\
129	0.843031666666667\\
130	0.84311\\
131	0.843147777777778\\
132	0.843226666666667\\
133	0.843497222222222\\
134	0.843858888888889\\
135	0.843957777777778\\
136	0.844367222222222\\
137	0.844258888888889\\
138	0.844375555555555\\
139	0.844377777777778\\
140	0.844378888888889\\
141	0.844669444444444\\
142	0.844918333333333\\
143	0.84497\\
144	0.845066666666667\\
145	0.845383888888889\\
146	0.845455555555556\\
147	0.845477777777777\\
148	0.845666111111111\\
149	0.845782777777778\\
150	0.846027777777778\\
151	0.846148333333333\\
152	0.846431111111111\\
153	0.846578888888889\\
154	0.846834444444445\\
155	0.846881111111111\\
156	0.847058888888889\\
157	0.847219444444445\\
158	0.847267222222222\\
159	0.847297222222222\\
160	0.847527222222222\\
161	0.847402777777778\\
162	0.847510555555555\\
163	0.847397777777778\\
164	0.847543888888889\\
165	0.847661111111111\\
166	0.84771\\
167	0.847788333333333\\
168	0.848022222222222\\
169	0.848197222222222\\
170	0.848342777777778\\
171	0.848332222222222\\
172	0.848657777777778\\
173	0.848681111111111\\
174	0.848847777777778\\
175	0.848912777777778\\
176	0.848925\\
177	0.848918888888889\\
178	0.848987222222222\\
179	0.848976111111111\\
180	0.848959444444445\\
181	0.849155555555556\\
182	0.849285\\
183	0.849415555555556\\
184	0.849298888888889\\
185	0.849297222222222\\
186	0.84951\\
187	0.849595\\
188	0.849741111111111\\
189	0.849841111111111\\
190	0.850125\\
191	0.850082777777778\\
192	0.850166111111111\\
193	0.850211111111111\\
194	0.850299444444444\\
195	0.850444444444444\\
196	0.850356666666667\\
197	0.85052\\
198	0.850736111111111\\
199	0.850870555555556\\
200	0.850775555555556\\
201	0.851015555555556\\
202	0.851004444444444\\
203	0.851092222222222\\
204	0.851165555555556\\
205	0.851242222222222\\
206	0.851374444444445\\
207	0.851551111111111\\
208	0.851533333333333\\
209	0.851608888888889\\
210	0.851706111111111\\
211	0.851785555555555\\
212	0.851893888888889\\
213	0.852106666666667\\
214	0.852200555555556\\
215	0.852235555555556\\
216	0.852399444444444\\
217	0.852376111111111\\
218	0.852415\\
219	0.852385\\
220	0.852555555555556\\
221	0.85261\\
222	0.852741666666667\\
223	0.852831111111111\\
224	0.852969444444444\\
225	0.853129444444444\\
226	0.853130555555556\\
227	0.85309\\
228	0.85313\\
229	0.853165555555556\\
230	0.853171111111111\\
231	0.853226666666667\\
232	0.853368888888889\\
233	0.853497777777778\\
234	0.853580555555556\\
235	0.853688333333334\\
236	0.853777777777778\\
237	0.85374\\
238	0.853809444444445\\
239	0.853799444444444\\
240	0.853915555555556\\
241	0.854018333333333\\
242	0.85405\\
243	0.854172777777778\\
244	0.85429\\
245	0.854346666666667\\
246	0.854368888888889\\
247	0.854506666666667\\
248	0.854607777777778\\
249	0.854736111111111\\
250	0.854782222222222\\
251	0.854795555555555\\
};
\addlegendentry{Suburban}

\addplot [color=mycolor4, line width=3.0pt, mark repeat={20}, mark size=6.0pt, mark=o, mark options={solid, mycolor4}]
  table[row sep=crcr]{%
1	0.103756666666667\\
2	0.166229444444444\\
3	0.25398\\
4	0.327144\\
5	0.392924\\
6	0.438730333333333\\
7	0.472944\\
8	0.498407333333333\\
9	0.520379\\
10	0.549556\\
11	0.57679\\
12	0.599744\\
13	0.624868\\
14	0.644236333333333\\
15	0.659599\\
16	0.674985\\
17	0.688547\\
18	0.697370333333333\\
19	0.708933333333333\\
20	0.717392\\
21	0.724759333333333\\
22	0.730099\\
23	0.734057666666667\\
24	0.734864666666667\\
25	0.737815\\
26	0.738301\\
27	0.741396\\
28	0.744452\\
29	0.745840666666667\\
30	0.748896\\
31	0.753064333333333\\
32	0.754511\\
33	0.758283333333334\\
34	0.763375333333334\\
35	0.764602666666667\\
36	0.766547333333333\\
37	0.769959666666667\\
38	0.772334666666667\\
39	0.773351666666667\\
40	0.776351\\
41	0.778089\\
42	0.780006333333333\\
43	0.780806333333333\\
44	0.784789666666667\\
45	0.785951333333333\\
46	0.786549666666667\\
47	0.785702333333333\\
48	0.786597333333333\\
49	0.784862333333333\\
50	0.785101666666667\\
51	0.787147333333333\\
52	0.791127\\
53	0.79416\\
54	0.797487\\
55	0.800586666666667\\
56	0.801169333333334\\
57	0.800849666666667\\
58	0.799273333333333\\
59	0.799057\\
60	0.799609333333333\\
61	0.801178333333333\\
62	0.803146666666667\\
63	0.807065\\
64	0.8093\\
65	0.81025\\
66	0.811168666666667\\
67	0.81177\\
68	0.810846333333334\\
69	0.810196333333334\\
70	0.810531\\
71	0.810843666666667\\
72	0.810883\\
73	0.811426666666667\\
74	0.810971333333333\\
75	0.811516\\
76	0.812091333333333\\
77	0.811806333333334\\
78	0.812427\\
79	0.814686333333333\\
80	0.813149\\
81	0.812173666666667\\
82	0.813319\\
83	0.813549333333333\\
84	0.813922\\
85	0.817464666666667\\
86	0.819821333333333\\
87	0.820277333333333\\
88	0.821191666666667\\
89	0.821450333333334\\
90	0.821098\\
91	0.822060666666667\\
92	0.823591666666667\\
93	0.824687666666667\\
94	0.825893\\
95	0.826553333333333\\
96	0.827039333333333\\
97	0.826753333333333\\
98	0.826922333333333\\
99	0.826825333333333\\
100	0.826785333333333\\
101	0.825973666666667\\
102	0.825898666666667\\
103	0.825651\\
104	0.825443333333334\\
105	0.825810666666667\\
106	0.826628333333333\\
107	0.827016\\
108	0.826918333333333\\
109	0.826275\\
110	0.825638\\
111	0.824666\\
112	0.825094333333333\\
113	0.824734666666667\\
114	0.826260666666667\\
115	0.828780666666667\\
116	0.829744666666667\\
117	0.830667666666667\\
118	0.831350333333333\\
119	0.830035333333333\\
120	0.828117666666667\\
121	0.828399\\
122	0.827724\\
123	0.828517\\
124	0.829758\\
125	0.830847333333333\\
126	0.831417333333333\\
127	0.831689333333333\\
128	0.831964\\
129	0.832433\\
130	0.832410666666667\\
131	0.832368666666667\\
132	0.832872\\
133	0.833632\\
134	0.834123\\
135	0.834846\\
136	0.835475666666667\\
137	0.835695666666667\\
138	0.835781666666667\\
139	0.836155666666667\\
140	0.836698\\
141	0.837387\\
142	0.838110666666667\\
143	0.838603333333333\\
144	0.838782333333333\\
145	0.838702666666667\\
146	0.838695\\
147	0.838712666666667\\
148	0.838608666666667\\
149	0.838534\\
150	0.838748\\
151	0.838990666666667\\
152	0.839034\\
153	0.839084333333333\\
154	0.838947666666667\\
155	0.838832\\
156	0.838284333333333\\
157	0.838204333333333\\
158	0.838443333333333\\
159	0.839010333333333\\
160	0.839463\\
161	0.840140666666667\\
162	0.840616\\
163	0.840941333333333\\
164	0.840659333333333\\
165	0.840429666666667\\
166	0.840291\\
167	0.840264666666667\\
168	0.840271\\
169	0.840642\\
170	0.841151\\
171	0.841708333333333\\
172	0.841803333333333\\
173	0.841859333333333\\
174	0.842043666666667\\
175	0.84218\\
176	0.842348666666667\\
177	0.842834666666667\\
178	0.843186666666667\\
179	0.843427666666667\\
180	0.843385666666667\\
181	0.843427333333333\\
182	0.843511333333333\\
183	0.843446\\
184	0.843348\\
185	0.843346\\
186	0.843338666666667\\
187	0.843428333333333\\
188	0.843649\\
189	0.84372\\
190	0.843799666666667\\
191	0.843371666666667\\
192	0.842905\\
193	0.842828666666667\\
194	0.842837666666667\\
195	0.842927333333333\\
196	0.843529333333333\\
197	0.843931\\
198	0.844103333333333\\
199	0.844447333333333\\
200	0.843663333333333\\
201	0.84376\\
202	0.843401666666667\\
203	0.843299\\
204	0.843455666666667\\
205	0.844756666666667\\
206	0.845289\\
207	0.846504333333333\\
208	0.845625666666667\\
209	0.844562333333333\\
210	0.84352\\
211	0.842086\\
212	0.840510666666667\\
213	0.840911\\
214	0.84098\\
215	0.840967666666667\\
216	0.841037333333333\\
217	0.841120333333333\\
218	0.841195\\
219	0.841808666666667\\
220	0.842539666666667\\
221	0.843373\\
222	0.843990333333333\\
223	0.844443\\
224	0.844611333333333\\
225	0.844691333333333\\
226	0.843564\\
227	0.843978666666667\\
228	0.844058666666667\\
229	0.842467\\
230	0.843105333333333\\
231	0.843971666666667\\
232	0.844749666666666\\
233	0.844643666666667\\
234	0.847015\\
235	0.845585666666667\\
236	0.845554333333333\\
237	0.844572666666667\\
238	0.845214333333334\\
239	0.843563666666667\\
240	0.843596\\
241	0.843273666666667\\
242	0.842745333333333\\
243	0.841926666666667\\
244	0.843242333333333\\
245	0.844345666666667\\
246	0.845222333333333\\
247	0.846116\\
248	0.846941333333333\\
249	0.847154666666667\\
250	0.847495\\
251	0.847615\\
};
\addlegendentry{Urban}

\addplot [color=mycolor8, line width=3.0pt, mark repeat={20}, mark size=6.0pt, mark=o, mark options={solid, mycolor8}]
  table[row sep=crcr]{%
1	0.0957533333333334\\
2	0.119739444444444\\
3	0.165416111111111\\
4	0.192876111111111\\
5	0.243172777777778\\
6	0.250518333333333\\
7	0.259428333333333\\
8	0.30064\\
9	0.322875\\
10	0.331602222222222\\
11	0.342910555555555\\
12	0.350981666666667\\
13	0.369391666666667\\
14	0.395635555555556\\
15	0.390870555555556\\
16	0.397822222222222\\
17	0.393995555555556\\
18	0.401011666666667\\
19	0.422098333333333\\
20	0.458504444444444\\
21	0.466382222222222\\
22	0.487532222222222\\
23	0.493086111111111\\
24	0.479643888888889\\
25	0.473156666666667\\
26	0.49973\\
27	0.479134444444444\\
28	0.473123333333333\\
29	0.487052222222222\\
30	0.475847222222222\\
31	0.490471111111111\\
32	0.507282222222222\\
33	0.527237222222222\\
34	0.529297222222222\\
35	0.526573888888889\\
36	0.536455555555556\\
37	0.548753888888889\\
38	0.556926111111111\\
39	0.570308888888889\\
40	0.572765\\
41	0.560898333333333\\
42	0.551816111111111\\
43	0.556062222222222\\
44	0.558621111111111\\
45	0.571886666666667\\
46	0.580917222222222\\
47	0.594094444444444\\
48	0.601226111111111\\
49	0.603574444444444\\
50	0.61393\\
51	0.612042777777778\\
52	0.613281666666667\\
53	0.610008333333333\\
54	0.593888888888889\\
55	0.605715\\
56	0.606413333333333\\
57	0.621393888888889\\
58	0.629554444444444\\
59	0.632450555555556\\
60	0.619015\\
61	0.611650555555556\\
62	0.617442222222222\\
63	0.609165555555556\\
64	0.616184444444445\\
65	0.625027222222222\\
66	0.628875555555555\\
67	0.632635\\
68	0.638943888888889\\
69	0.638832777777778\\
70	0.643727777777778\\
71	0.633638888888889\\
72	0.635897222222222\\
73	0.643246111111111\\
74	0.646586666666667\\
75	0.655883888888889\\
76	0.650746111111111\\
77	0.657667222222222\\
78	0.661816111111111\\
79	0.667362777777778\\
80	0.665754444444444\\
81	0.665026666666667\\
82	0.667525555555556\\
83	0.643603333333333\\
84	0.654607777777778\\
85	0.644720555555556\\
86	0.649437222222222\\
87	0.663616666666667\\
88	0.663405\\
89	0.665583333333333\\
90	0.677249444444444\\
91	0.68132\\
92	0.682802222222222\\
93	0.681486111111111\\
94	0.686051111111111\\
95	0.691369444444444\\
96	0.695979444444444\\
97	0.678958333333333\\
98	0.674456666666667\\
99	0.680601666666667\\
100	0.686735\\
101	0.692455555555556\\
102	0.699091111111111\\
103	0.703688888888889\\
104	0.705086111111111\\
105	0.707190555555556\\
106	0.709828888888889\\
107	0.710885\\
108	0.712131666666667\\
109	0.711303333333333\\
110	0.693583333333333\\
111	0.688137777777778\\
112	0.691465555555556\\
113	0.693904444444445\\
114	0.699868888888889\\
115	0.700135\\
116	0.704472777777778\\
117	0.712380555555555\\
118	0.717698333333333\\
119	0.711518888888889\\
120	0.715092222222222\\
121	0.718998333333333\\
122	0.705211666666667\\
123	0.708317222222222\\
124	0.711375\\
125	0.712487777777778\\
126	0.717143888888889\\
127	0.717776666666667\\
128	0.72073\\
129	0.719678333333333\\
130	0.705690555555555\\
131	0.700832222222222\\
132	0.696266666666667\\
133	0.704898888888889\\
134	0.710538888888889\\
135	0.702886666666667\\
136	0.705135\\
137	0.704842777777778\\
138	0.708968888888889\\
139	0.706924444444444\\
140	0.709390555555556\\
141	0.714891111111111\\
142	0.716371666666667\\
143	0.71661\\
144	0.719572777777778\\
145	0.721198888888889\\
146	0.72624\\
147	0.728822222222222\\
148	0.731737222222222\\
149	0.733398888888889\\
150	0.731293888888889\\
151	0.733498888888889\\
152	0.737356111111111\\
153	0.738269444444444\\
154	0.736682222222222\\
155	0.739324444444444\\
156	0.705070555555556\\
157	0.709500555555556\\
158	0.716474444444444\\
159	0.720193888888889\\
160	0.724633888888889\\
161	0.725542222222222\\
162	0.722582777777778\\
163	0.726886111111111\\
164	0.730676111111111\\
165	0.734000555555555\\
166	0.738474444444444\\
167	0.74054\\
168	0.735653888888889\\
169	0.734865555555556\\
170	0.733308333333333\\
171	0.730535\\
172	0.729485\\
173	0.730878888888889\\
174	0.733972777777778\\
175	0.736682777777778\\
176	0.735193333333333\\
177	0.737681111111111\\
178	0.740465555555555\\
179	0.742895\\
180	0.736278888888889\\
181	0.73734\\
182	0.740645\\
183	0.741105555555556\\
184	0.742988333333333\\
185	0.740317777777778\\
186	0.742758888888889\\
187	0.745251666666667\\
188	0.746644444444444\\
189	0.747618333333333\\
190	0.747603888888889\\
191	0.749343888888889\\
192	0.743448888888889\\
193	0.746937777777778\\
194	0.747396666666667\\
195	0.749169444444444\\
196	0.749733333333333\\
197	0.749665\\
198	0.749237222222222\\
199	0.751621111111111\\
200	0.752152222222222\\
201	0.752304444444444\\
202	0.746563333333333\\
203	0.740821111111111\\
204	0.745072222222222\\
205	0.746371111111111\\
206	0.74875\\
207	0.751467777777778\\
208	0.752726111111111\\
209	0.754958333333333\\
210	0.737497777777778\\
211	0.741018888888889\\
212	0.744301111111111\\
213	0.746663888888889\\
214	0.748353333333333\\
215	0.748898888888889\\
216	0.750506666666667\\
217	0.753146111111111\\
218	0.754231111111111\\
219	0.757779444444445\\
220	0.760157777777778\\
221	0.761786666666667\\
222	0.762696666666667\\
223	0.759721666666667\\
224	0.753557777777778\\
225	0.755687222222222\\
226	0.758068333333333\\
227	0.760092222222222\\
228	0.760782777777778\\
229	0.763276666666667\\
230	0.763923333333333\\
231	0.766027777777778\\
232	0.766685555555556\\
233	0.768670555555556\\
234	0.768138888888889\\
235	0.770466666666667\\
236	0.771445\\
237	0.767810555555555\\
238	0.769909444444445\\
239	0.770886111111111\\
240	0.770631111111111\\
241	0.769775\\
242	0.771305\\
243	0.772797777777778\\
244	0.774233888888889\\
245	0.774993888888889\\
246	0.774677777777778\\
247	0.775575\\
248	0.773778333333334\\
249	0.775026666666667\\
250	0.776232222222222\\
251	0.77579\\
};
\addlegendentry{Dense urban}

\addplot [color=mycolor5, line width=3.0pt, mark repeat={20}, mark size=6.0pt, mark=o, mark options={solid, mycolor5}]
  table[row sep=crcr]{%
1	0.0970994444444444\\
2	0.112364444444444\\
3	0.155952222222222\\
4	0.151218888888889\\
5	0.154578333333333\\
6	0.166051111111111\\
7	0.163233333333333\\
8	0.176868888888889\\
9	0.190672777777778\\
10	0.18449\\
11	0.215425555555556\\
12	0.218123333333333\\
13	0.228007777777778\\
14	0.225352777777778\\
15	0.230346666666667\\
16	0.229031111111111\\
17	0.234615555555556\\
18	0.237400555555556\\
19	0.248759444444444\\
20	0.248255555555555\\
21	0.251399444444444\\
22	0.259046666666667\\
23	0.270792222222222\\
24	0.275772777777778\\
25	0.287898888888889\\
26	0.284847222222222\\
27	0.292421666666667\\
28	0.303313333333333\\
29	0.302073888888889\\
30	0.297289444444444\\
31	0.307972222222222\\
32	0.311291666666667\\
33	0.328872777777778\\
34	0.334883333333333\\
35	0.340573333333333\\
36	0.343906111111111\\
37	0.328325\\
38	0.328335\\
39	0.322869444444444\\
40	0.331282777777778\\
41	0.338051111111111\\
42	0.349368333333333\\
43	0.347493888888889\\
44	0.362903333333333\\
45	0.359284444444444\\
46	0.367088333333333\\
47	0.367253888888889\\
48	0.366807222222222\\
49	0.37677\\
50	0.384579444444445\\
51	0.395763888888889\\
52	0.390355555555555\\
53	0.39318\\
54	0.400178333333333\\
55	0.406162777777778\\
56	0.407634444444445\\
57	0.408656666666667\\
58	0.410661111111111\\
59	0.406729444444444\\
60	0.395704444444444\\
61	0.375614444444444\\
62	0.385860555555555\\
63	0.384804444444445\\
64	0.384799444444444\\
65	0.386095555555556\\
66	0.386674444444444\\
67	0.391651666666667\\
68	0.394671111111111\\
69	0.391436111111111\\
70	0.389531666666667\\
71	0.392491666666667\\
72	0.397767777777778\\
73	0.401637222222222\\
74	0.410428888888889\\
75	0.4142\\
76	0.420893888888889\\
77	0.42292\\
78	0.430991666666667\\
79	0.435507777777778\\
80	0.433733333333333\\
81	0.433388333333333\\
82	0.432867777777778\\
83	0.438616666666667\\
84	0.443882777777778\\
85	0.447193888888889\\
86	0.445991111111111\\
87	0.447782222222222\\
88	0.450746111111111\\
89	0.449517777777778\\
90	0.455969444444444\\
91	0.453010555555556\\
92	0.457641111111111\\
93	0.449671111111111\\
94	0.447651666666667\\
95	0.451726666666667\\
96	0.457677777777778\\
97	0.461872222222222\\
98	0.461853888888889\\
99	0.466871111111111\\
100	0.472195555555556\\
101	0.475042777777778\\
102	0.479048333333333\\
103	0.480186111111111\\
104	0.483952777777778\\
105	0.485601666666667\\
106	0.483193333333333\\
107	0.489691666666666\\
108	0.489110555555555\\
109	0.489425\\
110	0.492737777777778\\
111	0.497075\\
112	0.491809444444444\\
113	0.489997777777778\\
114	0.491368333333333\\
115	0.488590555555555\\
116	0.485005\\
117	0.486133888888889\\
118	0.488501111111111\\
119	0.488407777777778\\
120	0.489831111111111\\
121	0.491334444444444\\
122	0.491077222222222\\
123	0.494169444444444\\
124	0.498457222222222\\
125	0.493244444444444\\
126	0.495976666666667\\
127	0.491256666666667\\
128	0.497272222222222\\
129	0.499295\\
130	0.501856111111111\\
131	0.501336111111111\\
132	0.502207222222222\\
133	0.505251666666667\\
134	0.506683888888889\\
135	0.507691666666667\\
136	0.511734444444444\\
137	0.512802222222222\\
138	0.510923333333333\\
139	0.512655555555555\\
140	0.515175555555556\\
141	0.513373333333333\\
142	0.515387222222222\\
143	0.513816666666667\\
144	0.515615\\
145	0.520248333333333\\
146	0.520426111111111\\
147	0.518836111111111\\
148	0.517013333333333\\
149	0.518096111111111\\
150	0.520984444444445\\
151	0.521146666666667\\
152	0.524093888888889\\
153	0.526497777777778\\
154	0.528575555555556\\
155	0.526106666666666\\
156	0.528227777777778\\
157	0.530606666666667\\
158	0.532175\\
159	0.525395555555555\\
160	0.529114444444444\\
161	0.527873333333333\\
162	0.530808888888889\\
163	0.532048333333333\\
164	0.530567777777778\\
165	0.526465\\
166	0.528408333333333\\
167	0.528793333333333\\
168	0.531763333333333\\
169	0.533276111111111\\
170	0.533942222222222\\
171	0.536347777777778\\
172	0.540365555555556\\
173	0.534453333333333\\
174	0.537512222222222\\
175	0.539698888888889\\
176	0.541411111111111\\
177	0.544556666666667\\
178	0.543403888888889\\
179	0.539832777777778\\
180	0.543027777777778\\
181	0.544248888888889\\
182	0.538157222222222\\
183	0.536974444444445\\
184	0.534475\\
185	0.536791111111111\\
186	0.539494444444444\\
187	0.542726111111111\\
188	0.542292222222222\\
189	0.546055555555556\\
190	0.547862222222222\\
191	0.549059444444444\\
192	0.550992777777778\\
193	0.552577222222222\\
194	0.558308333333333\\
195	0.560050555555555\\
196	0.563254444444444\\
197	0.570222777777778\\
198	0.569187222222222\\
199	0.565406666666667\\
200	0.567425\\
201	0.568045555555555\\
202	0.567854444444444\\
203	0.563361666666667\\
204	0.560333888888889\\
205	0.561874444444444\\
206	0.564277222222222\\
207	0.566827222222222\\
208	0.567711111111111\\
209	0.569303888888889\\
210	0.571157777777778\\
211	0.572979444444444\\
212	0.575367777777778\\
213	0.576670555555555\\
214	0.579672222222222\\
215	0.581377777777778\\
216	0.581722777777778\\
217	0.583325\\
218	0.585816111111111\\
219	0.585717222222222\\
220	0.588354444444445\\
221	0.590464444444444\\
222	0.592106111111111\\
223	0.592581666666667\\
224	0.590941666666667\\
225	0.592045\\
226	0.591858333333333\\
227	0.59279\\
228	0.584534444444444\\
229	0.582738888888889\\
230	0.582636111111111\\
231	0.581188333333333\\
232	0.578898888888889\\
233	0.575395555555556\\
234	0.578527222222222\\
235	0.581975555555555\\
236	0.58314\\
237	0.583933333333333\\
238	0.585025555555556\\
239	0.582831111111111\\
240	0.585135555555555\\
241	0.587232222222222\\
242	0.589663333333333\\
243	0.587613333333333\\
244	0.589756666666667\\
245	0.589791111111111\\
246	0.59021\\
247	0.591225555555556\\
248	0.587572777777778\\
249	0.589138888888889\\
250	0.591156111111111\\
251	0.594593333333333\\
};
\addlegendentry{High-rise urban}

\end{axis}

\begin{axis}[%
width=6in,
height=4.8in,
at={(0in,0in)},
scale only axis,
xmin=0,
xmax=1,
ymin=0,
ymax=1,
axis line style={draw=none},
ticks=none,
axis x line*=bottom,
axis y line*=left
]
\end{axis}

%% file: Figures/F10_new.tex
%
%
\definecolor{mycolor1}{rgb}{0.00000,0.44700,0.74100}%
\definecolor{mycolor2}{rgb}{0.85000,0.32500,0.09800}%
\definecolor{mycolor3}{rgb}{0.92900,0.69400,0.12500}%
\definecolor{mycolor4}{rgb}{0.49400,0.18400,0.55600}%
\definecolor{mycolor5}{rgb}{0.74902,0.00000,0.74902}%
\definecolor{mycolor6}{rgb}{0.00000,0.44706,0.74118}%
\definecolor{mycolor7}{rgb}{0.63500,0.07800,0.18400}
\definecolor{mycolor8}{rgb}{0.46667,0.67451,0.18824}

\begin{axis}[%
width=8in,
height=7in,
at={(0.78in,0.556in)},
scale only axis,
xmin=0,
xmax=255,
xtick={0,50,...,250},
ytick={0,0.1,...,0.9},
xlabel style={font=\color{white!15!black}},
xlabel={\Huge{$\text{Communication rounds}$}},
ymin=0,
ymax=0.9,
ylabel style={font=\color{white!15!black}},
ylabel={\Huge{$\text{Training accuracy} $}},
axis background/.style={fill=white},
xmajorgrids,
ymajorgrids,
legend style={at={(0.503,0.538)}, anchor=south west, legend cell align=left, align=left, draw=white!15!black,font=\huge}
]
\addplot [color=mycolor1, line width=3.0pt, mark repeat={20}, mark size=6.0pt, mark=o, mark options={solid, mycolor1}]
  table[row sep=crcr]{%
1	0.0949022222222222\\
2	0.0953716666666667\\
3	0.101676111111111\\
4	0.0968205555555555\\
5	0.103165555555556\\
6	0.108489444444444\\
7	0.112282222222222\\
8	0.111786666666667\\
9	0.122089444444444\\
10	0.119932777777778\\
11	0.122867777777778\\
12	0.125033888888889\\
13	0.123455\\
14	0.131125555555556\\
15	0.130977222222222\\
16	0.133408333333333\\
17	0.136445555555556\\
18	0.137332777777778\\
19	0.143722222222222\\
20	0.149196666666667\\
21	0.148497777777778\\
22	0.148640555555556\\
23	0.157232777777778\\
24	0.159053888888889\\
25	0.151881666666667\\
26	0.144783333333333\\
27	0.148127777777778\\
28	0.151194444444444\\
29	0.144273888888889\\
30	0.146781666666667\\
31	0.147531666666667\\
32	0.153586111111111\\
33	0.15457\\
34	0.158715555555556\\
35	0.159678888888889\\
36	0.157658333333333\\
37	0.153586666666667\\
38	0.155268333333333\\
39	0.152830555555556\\
40	0.157759444444444\\
41	0.155482777777778\\
42	0.152095\\
43	0.145907222222222\\
44	0.13781\\
45	0.138223888888889\\
46	0.144507777777778\\
47	0.142438333333333\\
48	0.138010555555556\\
49	0.142106111111111\\
50	0.146229444444444\\
51	0.146308888888889\\
52	0.149037222222222\\
53	0.149356111111111\\
54	0.152393888888889\\
55	0.155057222222222\\
56	0.159737222222222\\
57	0.158851666666667\\
58	0.157764444444444\\
59	0.162377222222222\\
60	0.163191666666667\\
61	0.163120555555556\\
62	0.160814444444444\\
63	0.161723333333333\\
64	0.161551666666667\\
65	0.156623888888889\\
66	0.156236666666667\\
67	0.160117777777778\\
68	0.168252777777778\\
69	0.170295\\
70	0.167615555555556\\
71	0.165883333333333\\
72	0.169170555555556\\
73	0.168966666666667\\
74	0.169825\\
75	0.170366666666667\\
76	0.170185\\
77	0.169678333333333\\
78	0.167557777777778\\
79	0.171084444444444\\
80	0.165573333333333\\
81	0.167581111111111\\
82	0.174322777777778\\
83	0.176697777777778\\
84	0.177752777777778\\
85	0.173335\\
86	0.170426111111111\\
87	0.172256111111111\\
88	0.173471111111111\\
89	0.175248333333333\\
90	0.175797222222222\\
91	0.178822777777778\\
92	0.176057222222222\\
93	0.178010555555556\\
94	0.176000555555556\\
95	0.176442222222222\\
96	0.176905555555556\\
97	0.177723888888889\\
98	0.176730555555556\\
99	0.180143888888889\\
100	0.179391666666667\\
101	0.182467777777778\\
102	0.179526111111111\\
103	0.183031666666667\\
104	0.185762777777778\\
105	0.186807222222222\\
106	0.187737777777778\\
107	0.186370555555556\\
108	0.189631111111111\\
109	0.191942777777778\\
110	0.191662777777778\\
111	0.193460555555556\\
112	0.193273888888889\\
113	0.192775\\
114	0.194171666666667\\
115	0.194020555555556\\
116	0.195091111111111\\
117	0.197064444444444\\
118	0.194607777777778\\
119	0.195937777777778\\
120	0.197005555555556\\
121	0.200304444444444\\
122	0.201334444444444\\
123	0.205991666666667\\
124	0.207722222222222\\
125	0.209439444444444\\
126	0.210961111111111\\
127	0.211697777777778\\
128	0.209270555555556\\
129	0.210198333333333\\
130	0.211305\\
131	0.213291666666667\\
132	0.214877777777778\\
133	0.217618888888889\\
134	0.213258888888889\\
135	0.212701666666667\\
136	0.206807777777778\\
137	0.208843888888889\\
138	0.207652222222222\\
139	0.209040555555556\\
140	0.208726666666667\\
141	0.208505\\
142	0.210462222222222\\
143	0.211247222222222\\
144	0.211121666666667\\
145	0.211270555555556\\
146	0.210933333333333\\
147	0.213218333333333\\
148	0.213420555555556\\
149	0.217907777777778\\
150	0.218407222222222\\
151	0.217742777777778\\
152	0.216498333333333\\
153	0.215975\\
154	0.213860555555556\\
155	0.213653888888889\\
156	0.214426111111111\\
157	0.211668333333333\\
158	0.210742222222222\\
159	0.21244\\
160	0.212192222222222\\
161	0.212164444444444\\
162	0.215272222222222\\
163	0.213353888888889\\
164	0.214464444444444\\
165	0.214527222222222\\
166	0.215252222222222\\
167	0.21852\\
168	0.22192\\
169	0.221515555555556\\
170	0.223216666666667\\
171	0.223348888888889\\
172	0.225017222222222\\
173	0.220299444444444\\
174	0.218931111111111\\
175	0.219270555555556\\
176	0.221767777777778\\
177	0.222414444444444\\
178	0.222593888888889\\
179	0.223318888888889\\
180	0.224619444444444\\
181	0.227261111111111\\
182	0.226472222222222\\
183	0.227371666666667\\
184	0.22766\\
185	0.225588888888889\\
186	0.225261111111111\\
187	0.228098888888889\\
188	0.229134444444445\\
189	0.229655\\
190	0.230910555555556\\
191	0.231805\\
192	0.231452222222222\\
193	0.232409444444444\\
194	0.232790555555556\\
195	0.233451111111111\\
196	0.233735555555556\\
197	0.233506111111111\\
198	0.236026666666667\\
199	0.237882777777778\\
200	0.240155\\
201	0.238098333333333\\
202	0.238378333333333\\
203	0.235076666666667\\
204	0.236464444444444\\
205	0.236236666666667\\
206	0.241254444444444\\
207	0.241156666666667\\
208	0.241436111111111\\
209	0.243802777777778\\
210	0.240853888888889\\
211	0.234287777777778\\
212	0.234631111111111\\
213	0.232307777777778\\
214	0.233368888888889\\
215	0.237590555555556\\
216	0.237397222222222\\
217	0.237987222222222\\
218	0.238160555555556\\
219	0.238065\\
220	0.243077222222222\\
221	0.244332777777778\\
222	0.245082777777778\\
223	0.245065\\
224	0.245797777777778\\
225	0.24817\\
226	0.250169444444444\\
227	0.255877777777778\\
228	0.256707222222222\\
229	0.257988333333333\\
230	0.258711111111111\\
231	0.259713888888889\\
232	0.261055555555556\\
233	0.262957222222222\\
234	0.261986666666667\\
235	0.261548888888889\\
236	0.258456111111111\\
237	0.259271666666667\\
238	0.259253333333333\\
239	0.25955\\
240	0.258015\\
241	0.264690555555556\\
242	0.264888333333333\\
243	0.265262777777778\\
244	0.268210555555556\\
245	0.26941\\
246	0.268241666666667\\
247	0.26913\\
248	0.268695\\
249	0.26859\\
250	0.269709444444444\\
251	0.272175\\
};
\addlegendentry{Suburban}

\addplot [color=mycolor4, line width=3.0pt, mark repeat={20}, mark size=6.0pt, mark=o, mark options={solid, mycolor4}]
  table[row sep=crcr]{%
1	0.104283333333333\\
2	0.142285\\
3	0.242831666666667\\
4	0.284955\\
5	0.267181666666667\\
6	0.334915\\
7	0.399313333333333\\
8	0.428425\\
9	0.463308333333333\\
10	0.49606\\
11	0.461071666666667\\
12	0.479635\\
13	0.543108333333333\\
14	0.555646666666667\\
15	0.578123333333333\\
16	0.599316666666667\\
17	0.609723333333333\\
18	0.627885\\
19	0.62699\\
20	0.608743333333333\\
21	0.641388333333333\\
22	0.648415\\
23	0.659868333333333\\
24	0.656146666666667\\
25	0.627166666666667\\
26	0.644328333333333\\
27	0.684871666666667\\
28	0.688008333333333\\
29	0.678311666666667\\
30	0.699495\\
31	0.695335\\
32	0.691356666666667\\
33	0.701888333333333\\
34	0.704586666666667\\
35	0.712335\\
36	0.722351666666667\\
37	0.70725\\
38	0.72553\\
39	0.72491\\
40	0.737536666666667\\
41	0.729751666666667\\
42	0.708085\\
43	0.711725\\
44	0.721315\\
45	0.726616666666667\\
46	0.741403333333333\\
47	0.749583333333333\\
48	0.745626666666667\\
49	0.758706666666667\\
50	0.757428333333333\\
51	0.75301\\
52	0.744406666666667\\
53	0.746691666666667\\
54	0.751305\\
55	0.75029\\
56	0.752513333333333\\
57	0.763213333333333\\
58	0.761683333333333\\
59	0.760878333333333\\
60	0.76732\\
61	0.766181666666667\\
62	0.77049\\
63	0.771008333333334\\
64	0.773098333333333\\
65	0.776163333333333\\
66	0.771516666666667\\
67	0.780451666666667\\
68	0.771491666666667\\
69	0.77703\\
70	0.7793\\
71	0.78221\\
72	0.779068333333333\\
73	0.778831666666667\\
74	0.777546666666666\\
75	0.777668333333333\\
76	0.781043333333333\\
77	0.78593\\
78	0.780116666666667\\
79	0.789423333333333\\
80	0.788028333333333\\
81	0.790741666666667\\
82	0.792095\\
83	0.792151666666667\\
84	0.794433333333333\\
85	0.791298333333333\\
86	0.793258333333333\\
87	0.793298333333334\\
88	0.788795\\
89	0.790045\\
90	0.78755\\
91	0.790865\\
92	0.797281666666667\\
93	0.797656666666667\\
94	0.800428333333333\\
95	0.79873\\
96	0.799503333333333\\
97	0.796225\\
98	0.798353333333333\\
99	0.803166666666667\\
100	0.807911666666667\\
101	0.807048333333333\\
102	0.807638333333333\\
103	0.80753\\
104	0.806055\\
105	0.805116666666667\\
106	0.804418333333333\\
107	0.803131666666667\\
108	0.803741666666667\\
109	0.804408333333333\\
110	0.806585\\
111	0.807145\\
112	0.807196666666667\\
113	0.807466666666667\\
114	0.808165\\
115	0.807863333333333\\
116	0.807651666666667\\
117	0.80835\\
118	0.807921666666667\\
119	0.809235\\
120	0.80674\\
121	0.806785\\
122	0.806678333333333\\
123	0.805491666666667\\
124	0.809211666666667\\
125	0.808786666666667\\
126	0.809173333333333\\
127	0.811291666666667\\
128	0.810383333333334\\
129	0.811535\\
130	0.812785\\
131	0.814706666666667\\
132	0.814486666666667\\
133	0.814746666666667\\
134	0.816106666666667\\
135	0.81657\\
136	0.81539\\
137	0.815103333333333\\
138	0.813833333333333\\
139	0.81451\\
140	0.815931666666667\\
141	0.813733333333333\\
142	0.816151666666667\\
143	0.816941666666667\\
144	0.816546666666667\\
145	0.820191666666667\\
146	0.820193333333333\\
147	0.818483333333333\\
148	0.818338333333333\\
149	0.819233333333333\\
150	0.821168333333333\\
151	0.81979\\
152	0.819133333333333\\
153	0.819253333333333\\
154	0.81951\\
155	0.82092\\
156	0.820366666666667\\
157	0.819396666666667\\
158	0.820153333333333\\
159	0.821126666666667\\
160	0.821468333333333\\
161	0.818685\\
162	0.819518333333333\\
163	0.818741666666667\\
164	0.821105\\
165	0.821856666666667\\
166	0.822241666666667\\
167	0.821238333333334\\
168	0.821158333333334\\
169	0.820703333333334\\
170	0.822498333333333\\
171	0.822323333333333\\
172	0.822515\\
173	0.822741666666667\\
174	0.822653333333333\\
175	0.823345\\
176	0.823216666666667\\
177	0.822698333333334\\
178	0.823125\\
179	0.82364\\
180	0.824285\\
181	0.823676666666667\\
182	0.822745\\
183	0.82423\\
184	0.824488333333333\\
185	0.824743333333333\\
186	0.825715\\
187	0.826333333333334\\
188	0.827521666666667\\
189	0.826563333333333\\
190	0.826773333333333\\
191	0.826898333333334\\
192	0.827683333333334\\
193	0.825876666666667\\
194	0.826525\\
195	0.827511666666667\\
196	0.828085\\
197	0.82769\\
198	0.828948333333333\\
199	0.829588333333333\\
200	0.829326666666667\\
201	0.829741666666667\\
202	0.830686666666667\\
203	0.829793333333333\\
204	0.82928\\
205	0.828855\\
206	0.829131666666667\\
207	0.828256666666667\\
208	0.827351666666667\\
209	0.827646666666667\\
210	0.82641\\
211	0.828238333333333\\
212	0.830145\\
213	0.829473333333334\\
214	0.82914\\
215	0.828948333333333\\
216	0.830011666666667\\
217	0.828978333333333\\
218	0.827426666666667\\
219	0.829028333333333\\
220	0.828905\\
221	0.82881\\
222	0.828781666666667\\
223	0.8299\\
224	0.82986\\
225	0.83029\\
226	0.828808333333333\\
227	0.82954\\
228	0.830348333333333\\
229	0.830675\\
230	0.830625\\
231	0.83059\\
232	0.830536666666667\\
233	0.831071666666667\\
234	0.831566666666667\\
235	0.832191666666667\\
236	0.832\\
237	0.831998333333333\\
238	0.831833333333333\\
239	0.831443333333333\\
240	0.832695\\
241	0.833545\\
242	0.832753333333333\\
243	0.83332\\
244	0.833128333333333\\
245	0.832026666666667\\
246	0.831815\\
247	0.831836666666667\\
248	0.832576666666667\\
249	0.8331\\
250	0.833376666666667\\
251	0.832841666666667\\
};
\addlegendentry{Urban}

\addplot [color=mycolor8, line width=3.0pt, mark repeat={20}, mark size=6.0pt, mark=o, mark options={solid, mycolor8}]
  table[row sep=crcr]{%
1	0.100127777777778\\
2	0.132218333333333\\
3	0.256387777777778\\
4	0.315075\\
5	0.376237777777778\\
6	0.405836666666667\\
7	0.474966666666667\\
8	0.507049444444444\\
9	0.527743333333333\\
10	0.570661111111111\\
11	0.585972777777778\\
12	0.60117\\
13	0.617862777777778\\
14	0.642994444444444\\
15	0.649690555555556\\
16	0.664518333333333\\
17	0.664276666666667\\
18	0.677647222222222\\
19	0.685235555555556\\
20	0.694148888888889\\
21	0.699501666666667\\
22	0.704321666666667\\
23	0.708587222222222\\
24	0.716554444444444\\
25	0.722104444444445\\
26	0.732069444444444\\
27	0.736827222222222\\
28	0.737752777777777\\
29	0.736250555555556\\
30	0.742556111111111\\
31	0.743607222222222\\
32	0.744650555555555\\
33	0.745226666666667\\
34	0.747846111111111\\
35	0.754136111111111\\
36	0.756196111111111\\
37	0.757358333333334\\
38	0.759238333333333\\
39	0.759493333333333\\
40	0.764048333333333\\
41	0.769478333333333\\
42	0.772658888888889\\
43	0.774010555555556\\
44	0.776145\\
45	0.774553888888889\\
46	0.774918888888889\\
47	0.775919444444444\\
48	0.778495\\
49	0.78218\\
50	0.783067777777778\\
51	0.783672222222222\\
52	0.785962777777778\\
53	0.78662\\
54	0.788725555555556\\
55	0.789098888888889\\
56	0.792778333333333\\
57	0.79299\\
58	0.794965555555556\\
59	0.793340555555556\\
60	0.794178333333334\\
61	0.795114444444444\\
62	0.797418333333333\\
63	0.796396666666667\\
64	0.798111111111111\\
65	0.799531111111111\\
66	0.799963888888889\\
67	0.798298333333333\\
68	0.799198333333333\\
69	0.799411666666667\\
70	0.802277777777778\\
71	0.803058333333334\\
72	0.804918888888889\\
73	0.805615555555556\\
74	0.804358333333333\\
75	0.805518888888889\\
76	0.806442222222222\\
77	0.807992777777778\\
78	0.808348333333333\\
79	0.809872222222222\\
80	0.810773888888889\\
81	0.810757777777778\\
82	0.810240555555555\\
83	0.810727222222222\\
84	0.810133888888889\\
85	0.810501111111111\\
86	0.811355\\
87	0.811814444444445\\
88	0.812897222222222\\
89	0.812203333333333\\
90	0.813667777777778\\
91	0.814718888888889\\
92	0.814839444444445\\
93	0.814696666666666\\
94	0.814451111111111\\
95	0.81429\\
96	0.814564444444445\\
97	0.816383333333333\\
98	0.815416666666667\\
99	0.815465555555556\\
100	0.816438333333333\\
101	0.816031666666667\\
102	0.81751\\
103	0.817136666666667\\
104	0.818205555555556\\
105	0.818587777777778\\
106	0.820039444444444\\
107	0.819447222222222\\
108	0.818568888888889\\
109	0.819061111111111\\
110	0.818267777777778\\
111	0.818712777777778\\
112	0.819738888888889\\
113	0.82041\\
114	0.819806111111111\\
115	0.820559444444444\\
116	0.821274444444444\\
117	0.820848333333333\\
118	0.822298888888889\\
119	0.822182777777778\\
120	0.821573333333333\\
121	0.822199444444444\\
122	0.822861666666666\\
123	0.8226\\
124	0.823166111111111\\
125	0.824280555555556\\
126	0.823983888888889\\
127	0.824617777777778\\
128	0.825529444444444\\
129	0.825178888888889\\
130	0.825910555555555\\
131	0.826465\\
132	0.825993888888889\\
133	0.826541666666666\\
134	0.826338888888889\\
135	0.825762222222222\\
136	0.82617\\
137	0.827488888888889\\
138	0.827230555555556\\
139	0.827225555555556\\
140	0.827789444444444\\
141	0.82799\\
142	0.827877222222222\\
143	0.829065\\
144	0.829723888888889\\
145	0.829412777777778\\
146	0.829721666666666\\
147	0.829847777777778\\
148	0.830444444444444\\
149	0.830405\\
150	0.830619444444444\\
151	0.830285\\
152	0.830588333333333\\
153	0.830414444444444\\
154	0.831993888888889\\
155	0.832001666666666\\
156	0.832154444444444\\
157	0.832471666666667\\
158	0.833201111111111\\
159	0.833412222222222\\
160	0.833056666666666\\
161	0.833246666666666\\
162	0.833173888888889\\
163	0.83347\\
164	0.833356666666667\\
165	0.834037222222223\\
166	0.834077777777778\\
167	0.834422222222222\\
168	0.83459\\
169	0.834156666666667\\
170	0.834006666666667\\
171	0.834359444444444\\
172	0.834541666666667\\
173	0.835313333333333\\
174	0.835405\\
175	0.835176111111111\\
176	0.835593333333333\\
177	0.835042777777778\\
178	0.83545\\
179	0.835185555555556\\
180	0.835132777777778\\
181	0.834686111111111\\
182	0.834903333333333\\
183	0.834572777777778\\
184	0.834751111111111\\
185	0.834950555555556\\
186	0.835592222222222\\
187	0.83595\\
188	0.836212222222222\\
189	0.836068888888889\\
190	0.836427777777778\\
191	0.836540555555555\\
192	0.836810555555555\\
193	0.837116111111111\\
194	0.837193888888889\\
195	0.837345\\
196	0.837303888888889\\
197	0.837312222222222\\
198	0.837711111111111\\
199	0.837896666666667\\
200	0.838496666666667\\
201	0.83854\\
202	0.838418333333333\\
203	0.838966666666667\\
204	0.838901666666667\\
205	0.838677777777778\\
206	0.838592777777778\\
207	0.838732777777778\\
208	0.838937222222222\\
209	0.839136111111111\\
210	0.839288888888889\\
211	0.839086111111111\\
212	0.839458888888889\\
213	0.839281666666666\\
214	0.839325555555555\\
215	0.83921\\
216	0.839005555555556\\
217	0.839013888888889\\
218	0.839404444444444\\
219	0.839688333333333\\
220	0.839483333333333\\
221	0.839586111111111\\
222	0.840008333333333\\
223	0.840163333333333\\
224	0.840654444444444\\
225	0.840600555555555\\
226	0.840925555555555\\
227	0.841201666666667\\
228	0.841085555555555\\
229	0.841147777777778\\
230	0.841402222222222\\
231	0.841251666666667\\
232	0.841243888888889\\
233	0.841146666666667\\
234	0.841048888888889\\
235	0.841427222222222\\
236	0.84166\\
237	0.841713888888889\\
238	0.842091111111111\\
239	0.842443333333333\\
240	0.842125\\
241	0.842286666666667\\
242	0.842033888888889\\
243	0.841962777777778\\
244	0.842085\\
245	0.842211111111111\\
246	0.842354444444445\\
247	0.842478888888889\\
248	0.842685\\
249	0.842537777777778\\
250	0.842598888888889\\
251	0.842547777777778\\
};
\addlegendentry{Dense urban}

\addplot [color=mycolor5, line width=3.0pt, mark repeat={20}, mark size=6.0pt, mark=o, mark options={solid, mycolor5}]
table[row sep=crcr]{%
1	0.0957138888888889\\
2	0.156571666666667\\
3	0.349812222222222\\
4	0.406320555555556\\
5	0.471947777777778\\
6	0.509734444444444\\
7	0.548872222222222\\
8	0.586694444444445\\
9	0.629444444444445\\
10	0.649485555555556\\
11	0.666653888888889\\
12	0.684099444444444\\
13	0.693082222222222\\
14	0.709405\\
15	0.711675\\
16	0.714338888888889\\
17	0.716001111111111\\
18	0.730911111111111\\
19	0.738544444444444\\
20	0.743785\\
21	0.745608333333333\\
22	0.752116666666667\\
23	0.758071111111111\\
24	0.760848888888889\\
25	0.765966666666667\\
26	0.768434444444445\\
27	0.76852\\
28	0.772104444444445\\
29	0.775398333333333\\
30	0.77492\\
31	0.777874444444444\\
32	0.78232\\
33	0.785786111111111\\
34	0.785327222222222\\
35	0.786058333333333\\
36	0.789311111111111\\
37	0.791690555555556\\
38	0.795502777777778\\
39	0.794722222222222\\
40	0.799648888888889\\
41	0.800578888888889\\
42	0.801555\\
43	0.801337222222222\\
44	0.804512222222222\\
45	0.805976111111111\\
46	0.806664444444444\\
47	0.806461111111111\\
48	0.807067222222222\\
49	0.80686\\
50	0.808077777777778\\
51	0.809048333333333\\
52	0.809031111111111\\
53	0.808325555555555\\
54	0.808648333333333\\
55	0.810816111111111\\
56	0.810969444444444\\
57	0.812637222222222\\
58	0.814137222222222\\
59	0.814845555555556\\
60	0.814265555555555\\
61	0.816161111111111\\
62	0.815792777777778\\
63	0.815547777777778\\
64	0.815938333333333\\
65	0.817571111111111\\
66	0.818339444444444\\
67	0.818738888888889\\
68	0.819485555555556\\
69	0.820296666666667\\
70	0.821600555555555\\
71	0.822141111111111\\
72	0.821963333333333\\
73	0.822371666666667\\
74	0.822506111111111\\
75	0.822836666666667\\
76	0.822696666666667\\
77	0.822700555555555\\
78	0.823278888888889\\
79	0.823538333333333\\
80	0.824833888888889\\
81	0.824721111111111\\
82	0.826050555555556\\
83	0.825555\\
84	0.826875555555555\\
85	0.826283333333333\\
86	0.827125\\
87	0.827621666666667\\
88	0.828455555555556\\
89	0.828948888888889\\
90	0.828787777777778\\
91	0.829396111111111\\
92	0.829386111111111\\
93	0.829421666666667\\
94	0.830189444444444\\
95	0.830511666666667\\
96	0.831102222222222\\
97	0.831412777777778\\
98	0.831742777777778\\
99	0.831507777777778\\
100	0.832272222222222\\
101	0.832258333333333\\
102	0.832342222222222\\
103	0.832568333333333\\
104	0.832487222222222\\
105	0.832944444444445\\
106	0.833382222222222\\
107	0.833182222222222\\
108	0.833773333333333\\
109	0.834250555555555\\
110	0.834161111111111\\
111	0.834803888888889\\
112	0.834908333333333\\
113	0.83538\\
114	0.835965\\
115	0.836593888888889\\
116	0.836430555555555\\
117	0.836866666666667\\
118	0.837015555555556\\
119	0.837256111111111\\
120	0.837051666666666\\
121	0.837408888888888\\
122	0.837543888888889\\
123	0.837678888888889\\
124	0.837554444444444\\
125	0.838065\\
126	0.838043333333333\\
127	0.838251111111111\\
128	0.838468888888889\\
129	0.838908888888889\\
130	0.838973333333333\\
131	0.839222777777778\\
132	0.839513888888889\\
133	0.839725\\
134	0.839554444444444\\
135	0.839485555555555\\
136	0.839441666666667\\
137	0.839490555555556\\
138	0.839533888888889\\
139	0.839833333333333\\
140	0.840119444444445\\
141	0.840427222222222\\
142	0.840245555555555\\
143	0.840467777777778\\
144	0.840889444444444\\
145	0.841126666666666\\
146	0.841241666666666\\
147	0.841293888888889\\
148	0.841150555555556\\
149	0.841333333333333\\
150	0.841407222222223\\
151	0.841260555555556\\
152	0.841732777777778\\
153	0.841792777777778\\
154	0.842036666666667\\
155	0.841916666666667\\
156	0.842259444444444\\
157	0.842493888888889\\
158	0.842805\\
159	0.843091111111111\\
160	0.843390555555556\\
161	0.843519444444444\\
162	0.843710555555556\\
163	0.843653333333333\\
164	0.843436666666667\\
165	0.843789444444445\\
166	0.844108333333333\\
167	0.84413\\
168	0.844108888888889\\
169	0.844143333333333\\
170	0.844393888888889\\
171	0.844429444444444\\
172	0.844343888888889\\
173	0.844623888888889\\
174	0.844997222222222\\
175	0.845142777777778\\
176	0.845378333333333\\
177	0.845257222222222\\
178	0.845155\\
179	0.845512777777778\\
180	0.845675\\
181	0.845868888888889\\
182	0.845935\\
183	0.846019444444444\\
184	0.846355555555556\\
185	0.846697777777778\\
186	0.847005555555555\\
187	0.846838333333334\\
188	0.846873888888889\\
189	0.846999444444444\\
190	0.847203333333333\\
191	0.847082222222222\\
192	0.847185555555556\\
193	0.847213888888889\\
194	0.847253888888889\\
195	0.84716\\
196	0.847447222222222\\
197	0.847831111111111\\
198	0.848001666666667\\
199	0.848332777777778\\
200	0.848407777777778\\
201	0.848347222222222\\
202	0.848231111111111\\
203	0.848251111111111\\
204	0.848375\\
205	0.848486111111111\\
206	0.848788333333333\\
207	0.848971666666667\\
208	0.849098333333334\\
209	0.849117777777778\\
210	0.84928\\
211	0.849417222222222\\
212	0.849173888888889\\
213	0.849282222222222\\
214	0.849370555555555\\
215	0.849413333333333\\
216	0.849288888888889\\
217	0.849543888888889\\
218	0.849668888888889\\
219	0.849938333333333\\
220	0.849818888888889\\
221	0.850021666666667\\
222	0.850016111111111\\
223	0.849963333333333\\
224	0.850115555555555\\
225	0.850078888888889\\
226	0.850025\\
227	0.850245\\
228	0.85048\\
229	0.850506111111111\\
230	0.850521666666667\\
231	0.850596111111111\\
232	0.850879444444445\\
233	0.850829444444445\\
234	0.850956666666667\\
235	0.85105\\
236	0.85118\\
237	0.851101111111111\\
238	0.851280555555556\\
239	0.851363888888889\\
240	0.851427777777778\\
241	0.851392222222222\\
242	0.851532222222222\\
243	0.851723333333333\\
244	0.851656111111111\\
245	0.851678333333333\\
246	0.851833888888889\\
247	0.851805555555555\\
248	0.851894444444444\\
249	0.851738333333333\\
250	0.851886111111111\\
251	0.852012222222223\\
};
\addlegendentry{High-rise urban}

\end{axis}

\begin{axis}[%
width=6in,
height=4.8in,
at={(0in,0in)},
scale only axis,
xmin=0,
xmax=1,
ymin=0,
ymax=1,
axis line style={draw=none},
ticks=none,
axis x line*=bottom,
axis y line*=left
]
\end{axis}